\documentclass[ALICE,manyauthors]{cernphprep}
\usepackage[comma,square,numbers,sort&compress]{natbib}
\usepackage{hyperref}
\usepackage{lineno}
\usepackage{xspace}
\usepackage[dvipsnames]{xcolor}
\usepackage[T1]{fontenc}
\usepackage{orcidlink}

\begin{document}
%

\newcommand{\pp}           {pp\xspace}
\newcommand{\ppbar}        {\mbox{$\mathrm {p\overline{p}}$}\xspace}
\newcommand{\XeXe}         {\mbox{Xe--Xe}\xspace}
\newcommand{\PbPb}         {\mbox{Pb--Pb}\xspace}
\newcommand{\pA}           {\mbox{pA}\xspace}
\newcommand{\pPb}          {\mbox{p--Pb}\xspace}
\newcommand{\AuAu}         {\mbox{Au--Au}\xspace}
\newcommand{\dAu}          {\mbox{d--Au}\xspace}

\newcommand{\s}            {\ensuremath{\sqrt{s}}\xspace}
\newcommand{\snn}          {\ensuremath{\sqrt{s_{\mathrm{NN}}}}\xspace}
\newcommand{\pt}           {\ensuremath{p_{\rm T}}\xspace}
\newcommand{\meanpt}       {$\langle p_{\mathrm{T}}\rangle$\xspace}
\newcommand{\ycms}         {\ensuremath{y_{\rm CMS}}\xspace}
\newcommand{\ylab}         {\ensuremath{y_{\rm lab}}\xspace}
\newcommand{\etarange}[1]  {\mbox{$\left | \eta \right |~<~#1$}}
\newcommand{\yrange}[1]    {\mbox{$\left | y \right |~<~#1$}}
\newcommand{\dndy}         {\ensuremath{\mathrm{d}N_\mathrm{ch}/\mathrm{d}y}\xspace}
\newcommand{\dndeta}       {\ensuremath{\mathrm{d}N_\mathrm{ch}/\mathrm{d}\eta}\xspace}
\newcommand{\avdndeta}     {\ensuremath{\langle\dndeta\rangle}\xspace}
\newcommand{\dNdy}         {\ensuremath{\mathrm{d}N_\mathrm{ch}/\mathrm{d}y}\xspace}
\newcommand{\Npart}        {\ensuremath{N_\mathrm{part}}\xspace}
\newcommand{\Ncoll}        {\ensuremath{N_\mathrm{coll}}\xspace}
\newcommand{\dEdx}         {\ensuremath{\textrm{d}E/\textrm{d}x}\xspace}
\newcommand{\RpPb}         {\ensuremath{R_{\rm pPb}}\xspace}

\newcommand{\nineH}        {$\sqrt{s}~=~0.9$~Te\kern-.1emV\xspace}
\newcommand{\seven}        {$\sqrt{s}~=~7$~Te\kern-.1emV\xspace}
\newcommand{\twoH}         {$\sqrt{s}~=~0.2$~Te\kern-.1emV\xspace}
\newcommand{\twosevensix}  {$\sqrt{s}~=~2.76$~Te\kern-.1emV\xspace}
\newcommand{\five}         {$\sqrt{s}~=~5.02$~Te\kern-.1emV\xspace}
\newcommand{\twosevensixnn}{$\sqrt{s_{\mathrm{NN}}}~=~2.76$~Te\kern-.1emV\xspace}
\newcommand{\fivenn}       {$\sqrt{s_{\mathrm{NN}}}~=~5.02$~Te\kern-.1emV\xspace}
\newcommand{\LT}           {L{\'e}vy-Tsallis\xspace}
\newcommand{\GeVc}         {Ge\kern-.1emV/$c$\xspace}
\newcommand{\MeVc}         {Me\kern-.1emV/$c$\xspace}
\newcommand{\TeV}          {Te\kern-.1emV\xspace}
\newcommand{\GeV}          {Ge\kern-.1emV\xspace}
\newcommand{\MeV}          {Me\kern-.1emV\xspace}
\newcommand{\GeVmass}      {Ge\kern-.2emV/$c^2$\xspace}
\newcommand{\MeVmass}      {Me\kern-.2emV/$c^2$\xspace}
\newcommand{\lumi}         {\ensuremath{\mathcal{L}}\xspace}
\newcommand{\GeVsqr}          {Ge\kern-.1emV$^2$\xspace}

\newcommand{\ITS}          {\rm{ITS}\xspace}
\newcommand{\TOF}          {\rm{TOF}\xspace}
\newcommand{\ZDC}          {\rm{ZDC}\xspace}
\newcommand{\ZDCs}         {\rm{ZDCs}\xspace}
\newcommand{\ZNA}          {\rm{ZNA}\xspace}
\newcommand{\ZNC}          {\rm{ZNC}\xspace}
\newcommand{\SPD}          {\rm{SPD}\xspace}
\newcommand{\SDD}          {\rm{SDD}\xspace}
\newcommand{\SSD}          {\rm{SSD}\xspace}
\newcommand{\TPC}          {\rm{TPC}\xspace}
\newcommand{\TRD}          {\rm{TRD}\xspace}
\newcommand{\VZERO}        {\rm{V0}\xspace}
\newcommand{\VZEROA}       {\rm{V0A}\xspace}
\newcommand{\VZEROC}       {\rm{V0C}\xspace}
\newcommand{\Vdecay} 	   {\ensuremath{V^{0}}\xspace}

\newcommand{\ee}           {\ensuremath{e^{+}e^{-}}} 
\newcommand{\pip}          {\ensuremath{\pi^{+}}\xspace}
\newcommand{\pim}          {\ensuremath{\pi^{-}}\xspace}
\newcommand{\kap}          {\ensuremath{\rm{K}^{+}}\xspace}
\newcommand{\kam}          {\ensuremath{\rm{K}^{-}}\xspace}
\newcommand{\pbar}         {\ensuremath{\rm\overline{p}}\xspace}
\newcommand{\kzero}        {\ensuremath{{\rm K}^{0}_{\rm{S}}}\xspace}
\newcommand{\lmb}          {\ensuremath{\Lambda}\xspace}
\newcommand{\almb}         {\ensuremath{\overline{\Lambda}}\xspace}
\newcommand{\Om}           {\ensuremath{\Omega^-}\xspace}
\newcommand{\Mo}           {\ensuremath{\overline{\Omega}^+}\xspace}
\newcommand{\X}            {\ensuremath{\Xi^-}\xspace}
\newcommand{\Ix}           {\ensuremath{\overline{\Xi}^+}\xspace}
\newcommand{\Xis}          {\ensuremath{\Xi^{\pm}}\xspace}
\newcommand{\Oms}          {\ensuremath{\Omega^{\pm}}\xspace}
\newcommand{\degree}       {\ensuremath{^{\rm o}}\xspace}


\newcommand{\Noon}         {\textbf{n$\mathbf{_O^O}$n}}
\newcommand{\Jpsi}{\ensuremath{{\rm J}/\psi}\xspace}
\newcommand{\psip}{\ensuremath{\psi(2S)}\xspace}
\newcommand{\Wgp}{\ensuremath{W_{\gamma \mathrm{p}}}\xspace}
\newcommand{\WgPb}{\ensuremath{W_{\gamma \mathrm{Pb,n}}}\xspace}
\newcommand{\WgAu}{\ensuremath{W_{\gamma \mathrm{Au,n}}}\xspace}
\newcommand{\ptdimuon}     {\ensuremath{p^{\mu\mu}_{\rm T}}\xspace}
\newcommand{\mdimuon}     {\ensuremath{m_{\mu\mu}}\xspace}
\newcommand{\pttwo}        {\ensuremath{p^{2}_{\rm T}}\xspace}
\newcommand{\mant}         {\ensuremath{|t|}\xspace}
\newcommand{\BR}           {\ensuremath{{\rm BR}(\Jpsi\to\mu^+\mu^-)}\xspace}
\newcommand{\axe}          {\ensuremath{{\rm Acc}\times\epsilon}\xspace}
\newcommand{\fC}           {\ensuremath{f_{\rm C}}\xspace}
\newcommand{\fD}           {\ensuremath{f_{\rm D}}\xspace}

\newcommand{\tbd}[1]{\colorbox{RoyalBlue}{\textcolor{white}{to be discussed:}}\textcolor{RoyalBlue}{~#1}}
\newcommand{\np}[1]{\colorbox{Blue}{\textcolor{white}{NatPhys:}}\textcolor{Blue}{~#1}}
\newcommand{\rojo}[1]{{\color{red}{#1}}}
\newcommand{\azul}[1]{{\color{blue}{#1}}}
\newcommand{\prl}[1]{\colorbox{Blue}{\textcolor{white}{PRL:}}\textcolor{Blue}{~#1}}

\begin{titlepage}
\PHyear{2025}       
\PHnumber{057}      
\PHdate{15 March}  

\title{
Evidence for \Jpsi suppression in incoherent photonuclear production 
}
\ShortTitle{Evidence for \Jpsi suppression in incoherent photonuclear production}   

\Collaboration{ALICE Collaboration\thanks{See Appendix~\ref{app:collab} for the list of collaboration members}}
\ShortAuthor{ALICE Collaboration} 

\begin{abstract}
According to quantum chromodynamics, at sufficiently high energy, the structure of hadrons reveals a dynamic equilibrium between gluon splitting and gluon recombination---a phenomenon known as saturation. 
The process of diffractive photonuclear production of a \Jpsi vector meson provides a direct insight into the gluon composition of hadrons. The \Jpsi production as a function of momentum transferred in the interaction, quantified by the Mandelstam-$t$ variable, serves as an excellent probe for studying the structure of hadrons within the impact-parameter plane, because different ranges in $t$ are sensitive to the dynamics of the gluon field at varying spatial size scales.
The ALICE collaboration has measured the energy dependence of incoherent photonuclear production of \Jpsi mesons off lead ions, at \fivenn, for three Mandelstam-$t$ intervals.
The energy dependence of the photonuclear cross section at the highest \mant range measured, $(0.81<\mant <1.44$) \GeVsqr, is sensitive to subnucleonic structures of the Pb target. The increase of the cross section with energy at large \mant shows evidence of suppression with respect to the increase seen at low \mant. The observed pattern of the energy evolution in data is similar to that of gluon saturation models. 
\end{abstract}

\end{titlepage}

\setcounter{page}{2} 


\section*{Introduction}
Deep-inelastic scattering experiments have demonstrated that the structure of protons in high-energy interactions is dominated by gluons, with their number increasing according to a power law with the interaction energy~\cite{H1:2015ubc}. Quantum chromodynamics (QCD), the prevailing theory of the strong force,
predicts that such high-energy conditions give rise to saturation: a dynamic equilibrium between splitting and recombination of gluons that plays a crucial role in taming the growth of the gluon number~\cite{Kovchegov:2012mbw}.
The search for saturation is currently one of the key research areas in perturbative QCD. It is, for example, one of the scientific pillars of the future electron--ion collider~\cite{Aschenauer:2017jsk}. Various observables have been put forward to seek experimental evidence of this phenomenon~\cite{Morreale:2021pnn}. However, to date, measurements potentially sensitive to saturation have also been explained by models that do not incorporate this effect. 

The diffractive photoproduction of the \Jpsi meson, where a photon fluctuates into a charm-anticharm pair that exchanges at least two gluons with the target before creating a \Jpsi meson, is a good tool for studying saturation owing to its strong sensitivity to the gluon density within the target~\cite{Ryskin:1992ui,Flett:2024htj}. 
This process has been extensively studied at the CERN Large Hadron Collider (LHC), employing protons and lead nuclei (Pb) as targets, see e.g. reviews~\cite{Contreras:2015dqa,Klein:2019qfb,ALICE:2022wpn} and Refs.~\cite{ALICE:2023jgu,CMS:2023snh,ALICE:2021tyx,ALICE:2023gcs,ALICE:2019tqa,ALICE:2023mfc,LHCb:2022ahs}.
The photons are provided by the electromagnetic field of the particles accelerated in the LHC, which can be represented as a flux of quasi-real photons. Photon-induced processes can be experimentally studied in ultra-peripheral collisions (UPCs), characterised by an impact parameter larger than the range of the strong interaction~\cite{Baltz:2007kq}.

In photonuclear production, the process is called coherent if the photon interacts with the full
nucleus, and incoherent if the photon interacts with a single nucleon in the target, which causes the
nucleus to break up. In both cases, the emitting nucleus remains intact. The incoherent processes are further classified into elastic, when the struck nucleon remains intact, and dissociative, when the target nucleon dissociates into a system with a mass larger than that of the nucleon.
By measuring the rapidity ($y$) and the transverse momentum ($\pt$) of the \Jpsi, the following kinematic variables can be determined event-by-event using $(\WgPb)^2 = m\snn\exp\left(-y\right)$ and $\mant=\pt^2$,
where $m$ is the invariant mass of the \Jpsi. The \Jpsi rapidity is measured in the laboratory frame with respect to the direction of the incoming Pb nucleus. \WgPb is the centre-of-mass energy per nucleon of the photon--lead system, $\snn$ is the centre-of-mass energy per nucleon pair of the nucleus--nucleus collision, and $t$ is the square of the four-momentum transfer in the interaction.

Saturation effects are expected to be more pronounced in heavy nuclei, such as Pb, and to manifest at lower energies compared to proton targets~\cite{McLerran:1993ni}. In photon--nucleus collisions, it is an established experimental observation that, at high energies, the gluon density in a nucleus composed of A nucleons is reduced compared to the expectation from A free nucleons~\cite{Armesto:2006ph}. 
Two primary theoretical frameworks have been proposed to explain this effect: saturation models, which describe non-linear dynamics of gluon densities, and shadowing models, which account for multiple scatterings of the probe in the nuclear colour field~\cite{Armesto:2006ph}. Both approaches have successfully described a variety of observables, leaving it unclear which of them better represents the underlying physics. Identifying observables capable of distinguishing between the predictions of saturation- and shadowing-based models is therefore crucial for advancing our understanding of these phenomena.

The energy dependence of the cross section of coherent \Jpsi photonuclear production, as a function of \WgPb, was recently studied at the LHC by the ALICE and CMS collaborations, across an energy range of $(20< \WgPb<633)$ \GeV~\cite{ALICE:2023jgu, CMS:2023snh}. There is also recent data from the STAR collaboration for $(15< \WgAu<25)$~\GeV~\cite{STAR:2023nos}.
Moreover, the ALICE collaboration has carried out measurements of the \mant dependence for both coherent and incoherent photonuclear production cross sections of \Jpsi mesons from lead nuclei at $\WgPb=125$~GeV~\cite{ALICE:2021tyx,ALICE:2023gcs}. Notably, the ALICE studies have highlighted the significance of the incoherent dissociative component for describing the data trend at \mant values near 1 GeV$^2$. The Mandelstam-$t$ variable is related to the impact parameter via a Fourier transform. This connection serves as an ideal tool for probing the gluon distribution across various size scales in the transverse plane of the target's trajectory, with larger (smaller) spatial scales prevailing at lower (higher) \mant values. Current data on coherent \Jpsi production above $\WgPb\sim 100$ GeV can be explained by saturation or shadowing models when their dependence on \WgPb or on \mant is studied separately.

In this Letter, the ALICE collaboration presents the first multi-differential measurement of the \WgPb and \mant dependence of the cross section for incoherent \Jpsi photonuclear production. The measurement is performed within the $(20 < \WgPb < 633)$ GeV range. The energy dependence is studied over three different \mant intervals, spanning from 0.09~GeV$^2$ to 1.44~GeV$^2$.

\section*{Analysis}
In the autumn of 2018, the LHC delivered Pb--Pb collisions at \fivenn. The products of the collisions were recorded with the ALICE experimental apparatus~\cite{ALICE:2008ngc,ALICE:2014sbx}. The selection of the events is composed of three sets of requirements. First, the \Jpsi meson is reconstructed from its decay into a $\mu^+\mu^-$ pair, with both muons measured either in ALICE’s central barrel at midrapidity ($|y| < 0.8$)~\cite{ALICE:2010tia,Alme:2010ke} or within the muon spectrometer~\cite{ALICE:1999aa} for muons at backward or forward rapidities ($2.5 < |y| < 4$). Second, the direction of the incoming lead nucleus that participates in the photonuclear interaction is determined by measuring neutrons, from the dissociation of the target nucleus, detected at beam rapidities using the zero-degree calorimeters (ZDCs)~\cite{Arnaldi:1999zz}. Last, the AD~\cite{Broz:2020ejr} and V0~\cite{ALICE:2013axi} detectors are employed to require vetoes on activity at other
rapidities than those of the muons and neutrons, thus selecting UPCs. AD, V0, and ZDC are two-armed systems with detectors placed at either side of the nominal interaction region along the direction of the incoming beams. The sides are labelled with the letters A and C, with the
muon spectrometer being located on side C. The two arms of AD, V0, and ZDC are thus denoted as ADA, ADC, V0A, V0C, ZNA, and ZNC. 
Cross sections are measured in three \mant intervals: $(0.09<\mant<0.36)$ \GeVsqr, $(0.36<\mant<0.81)$ \GeVsqr, and $(0.81<\mant<1.44)$ \GeVsqr. 
Four values of \WgPb are sampled for each \mant range: 
20~\GeV ($3.25<y<4$), 30~\GeV ($2.5<y<3.25$), 125~\GeV ($|y|<0.8$), and 633~\GeV ($-4<y<-2.5$). The quoted values of \WgPb correspond to the middle of the rapidity range used for the measurement.
 For the data at $\WgPb=125$ \GeV, the dimuon pair is measured with the central barrel detectors, and measurements at the other energies utilise the muon spectrometer to reconstruct the \Jpsi. At $\WgPb=633$~\GeV, the neutrons are detected with ZNC, 
 while at $\WgPb=20$ GeV and $\WgPb=30$ GeV with ZNA.
 
This work builds upon established techniques and methodologies previously developed by the ALICE collaboration, as demonstrated in recent analyses~\cite{ALICE:2023gcs,ALICE:2023jgu,ALICE:2023mfc}. The analysis for the \Jpsi mesons produced at midrapidity follows the same procedure and uses the same data set, of integrated luminosity $232\pm7$ $\mu$b$^{-1}$, as those described in Ref.~\cite{ALICE:2023gcs}.
The analysis of muon pairs reconstructed with the muon spectrometer follows the same technique employed in Ref.~\cite{ALICE:2023jgu}. Since the \mant dependence has never been measured before at forward rapidity, details of the analysis are given here. Such events are selected based on a trigger requiring two muons of opposite charge in the muon spectrometer and a veto on signals in the V0A detector. 
The cross section for incoherent photonuclear production of \Jpsi is given by
\begin{equation}
\frac{d\sigma_{{\rm Pb}\rightarrow \Jpsi {\rm Pb}}}{d\mant}(y)= \frac{1}{n_{\gamma}(y)}\frac{d^2\sigma^{\Jpsi}_{\rm Pb Pb}}{d\mant dy}= 
\frac{1}{n_\gamma}
\frac{N_{\Jpsi}/(1+\fC+\fD)}{(\axe)\cdot\BR\cdot\mathcal{L}
\cdot\Delta y\cdot \Delta\mant}.
\end{equation}
Here, $N_{\Jpsi}$ is the \Jpsi raw yield, \fC and \fD are factors to account for the contamination by coherent \Jpsi production and feed down from coherent and incoherent \psip, $\Delta y$ ($\Delta\mant$) accounts for the width of the rapidity (\mant) region where the measurement is performed, and ($\axe$)
 accounts for acceptance and efficiency in the muon spectrometer for the incoherent \Jpsi process, the neutron detection efficiency in the ZDCs 
 as well as the effect of pile-up on the efficiency of AD, V0, and ZDC vetoes.
 The branching ratio of the dimuon decay of the \Jpsi, \BR, is $0.05961\pm0.00033$~\cite{ParticleDataGroup:2024cfk}. The integrated luminosity, $\mathcal{L}$, for the muon-spectrometer sample is $(533\pm13)\ \mu{\rm b}^{-1}$, determined using reference cross sections measured in van der Meer scans~\cite{ALICE:2022xir}. The photon flux, $n_\gamma$, is calculated with the \Noon~Monte Carlo program~\cite{Broz:2019kpl}. 

The \Jpsi yield is extracted from unbinned extended log-likelihood fits to the invariant mass distributions of $\mu^+\mu^-$ pairs (see supplementary material). 
The \Jpsi and \psip contributions are modelled with double-sided Crystal Ball distributions~\cite{ALICE:2015qse}. The free parameters are the normalisation of both distributions and the pole-mass value of the \Jpsi. The rest of the parameters are fixed as explained in Ref.~\cite{ALICE:2023jgu}.
 The contribution from non-resonant $\mu^+\mu^-$ pairs is modelled by a decaying exponential distribution with the slope and normalisation being free parameters. 
The \psip contribution is not visible at high \WgPb. In the two largest \mant
ranges at the largest \WgPb, the fits are consistent with no background.
As the resolution is substantially smaller than the width of the \mant range, migration effects play no significant role.
 The \Jpsi yield varies from about 1000 candidates at the lowest \WgPb and \mant, to about 40 candidates at the largest \WgPb and \mant.
 
Both \fC and \fD are determined from a fit to the transverse momentum distribution of the \Jpsi yield in the invariant mass range $(2.85,3.35)$ \GeVmass (see supplementary material).
The fit model consists of four templates obtained from Monte Carlo simulated data, and one distribution function. A binned likelihood fit is used. 
The four Monte Carlo templates, from the STARlight Monte Carlo program~\cite{Klein:2016yzr}, correspond to the following processes: coherent and elastic incoherent production of either \Jpsi
or $\psip$ that after decaying produces a \Jpsi. The distribution for the contribution of the incoherent dissociative production of \Jpsi is based on the H1 parameterisation of this process~\cite{H1:2013okq}. This model has two parameters, $n$ and $b$, describing the power-law behaviour at large \mant and the exponential shape at low \mant, respectively. These two parameters are strongly correlated, so the first is fixed to $n=3.58$, as found by the H1 collaboration, and the second is a free parameter. The contamination from coherent photonuclear production of \Jpsi, \fC, is a few per mill or less for all cases, except for the lowest \mant at the highest \WgPb, where it is a 10\% effect. Similarly, in this region, the contamination from feed-down from \psip is significant, reaching 20\%. Otherwise, \fD ranges from a few per cent to a few per mill level. 

The acceptance and efficiency for detecting both muons from the \Jpsi decay are obtained using a reweighted STARlight Monte Carlo. This reweighting accounts for the absence of the incoherent dissociative process in the generator. The procedure is applied as a function of transverse momentum in each rapidity region to ensure that the reweighted MC accurately matches the raw data. The resulting correction varies from 0.10 to 0.13 with \WgPb and remains fairly constant with \mant.
The efficiency of the ZDCs to detect neutrons is 0.95, where this number is determined using the energy spectra measured in the calorimeters and the Poisson model of Ref.~\cite{Dmitrieva:2018rpi}. 
Pile-up has two contributions: signals from independent interactions, and signals from the same pair of Pb ions having two independent photon exchanges. For ZNA and ZNC the former amounts to 2.4\%, while the latter is 5\% at low \WgPb and 30\% at large \WgPb. Both types of pile-up are evaluated simultaneously and amount to 6\% and 20\% at low and large \WgPb, respectively. At large \WgPb, interactions occur at smaller impact parameters than those at low \WgPb. At small impact parameters the probability of extra independent photon exchanges is larger, which explains the difference in the quoted values~\cite{Broz:2019kpl}. 

The uncertainty on the signal extraction is determined by varying the value of the tail parameters of the Crystal Ball distribution, the invariant-mass range used for the fit, and using a first order polynomial as the model for non-resonant background. This uncertainty varies from a few per mill to up to 3.5\%. It is uncorrelated across \WgPb and \mant. The uncertainty on the measurement of the muon tracks is described in detail in Ref.~\cite{ALICE:2023jgu}. Its value is 6.9\%, dominated by the uncertainty on the muon trigger.
The uncertainty on \fC and \fD is given by the fit used to obtain these parameters. It is 6\% (\fC) and 5\% (\fD) for the lowest \mant region at the largest \WgPb. 
The uncertainty on \axe derived from the reweighted STARlight MC is well below 1\% and therefore considered negligible for this analysis. 
The uncertainty on vetoing activity in V0C is 5.6\%, obtained by varying the amount of accepted activity in this detector. The uncertainty on vetoing activity in V0A is 5\% and originates from the knowledge of the V0A trigger efficiency as determined from data.
The uncertainty on the efficiency of the ZDCs is 1\%, and arises from the modelling of the energy spectra in the calorimeters. The uncertainty on the same-Pb-pair pile-up is negligible at low \WgPb (4 per mill level), and 3\% at large \WgPb. This uncertainty is related to the size of the data sample used for its determination (see supplementary material). 
The uncertainty on the branching ratio, 0.6\%, is taken from the PDG~\cite{ParticleDataGroup:2024cfk}. The uncertainty on the luminosity is 2.5\%~\cite{ALICE:2022xir}. The uncertainty on the photon flux, 2\%, is obtained by varying the parameter of the nuclear radius in the Woods-Saxon distribution according to neutron-skin measurements~\cite{Abrahamyan:2012gp}.
Note that for 80\% of the events 
containing a dimuon with a mass around the \Jpsi mass, the dissociated system does not leave signals in the AD detector, which covers the pseudorapidity range $4.6\lesssim\eta\lesssim 6.9$. This is consistent with a boosted dissociated system with a steeply falling mass distribution. In this scenario the vetoes imposed with V0, located at more central pseudorapidities than AD, would reject at most a few percent of the cross section. Thus, no correction is considered for this effect.
The uncertainties on the signal extraction and on the measurement of the muon tracks are uncorrelated across \mant and \WgPb. All other uncertainties are considered as correlated.

\section*{Results}
\begin{figure}[!t]
\centering
\includegraphics[width=0.48\textwidth]{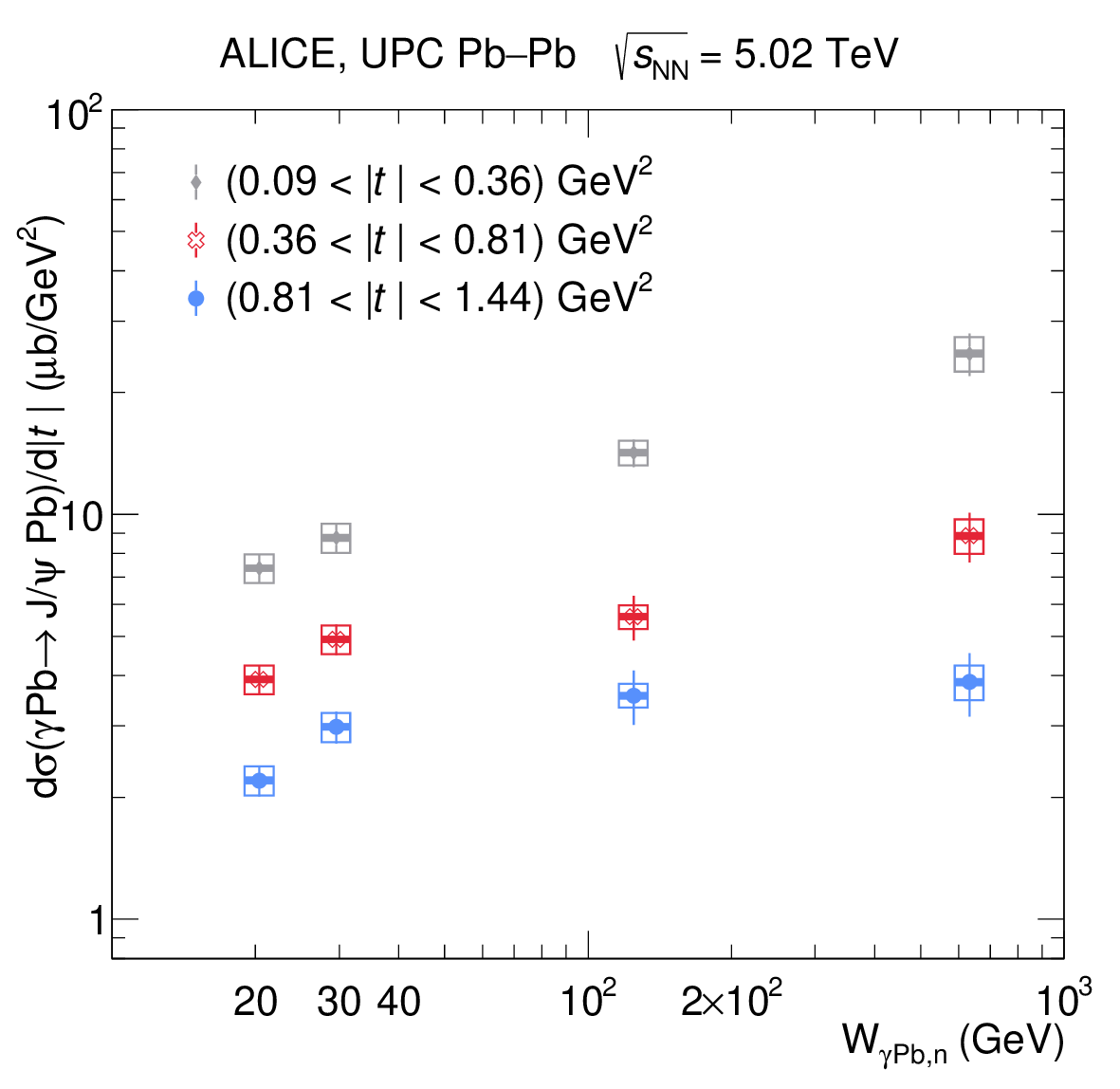}
\includegraphics[width=0.48\textwidth]{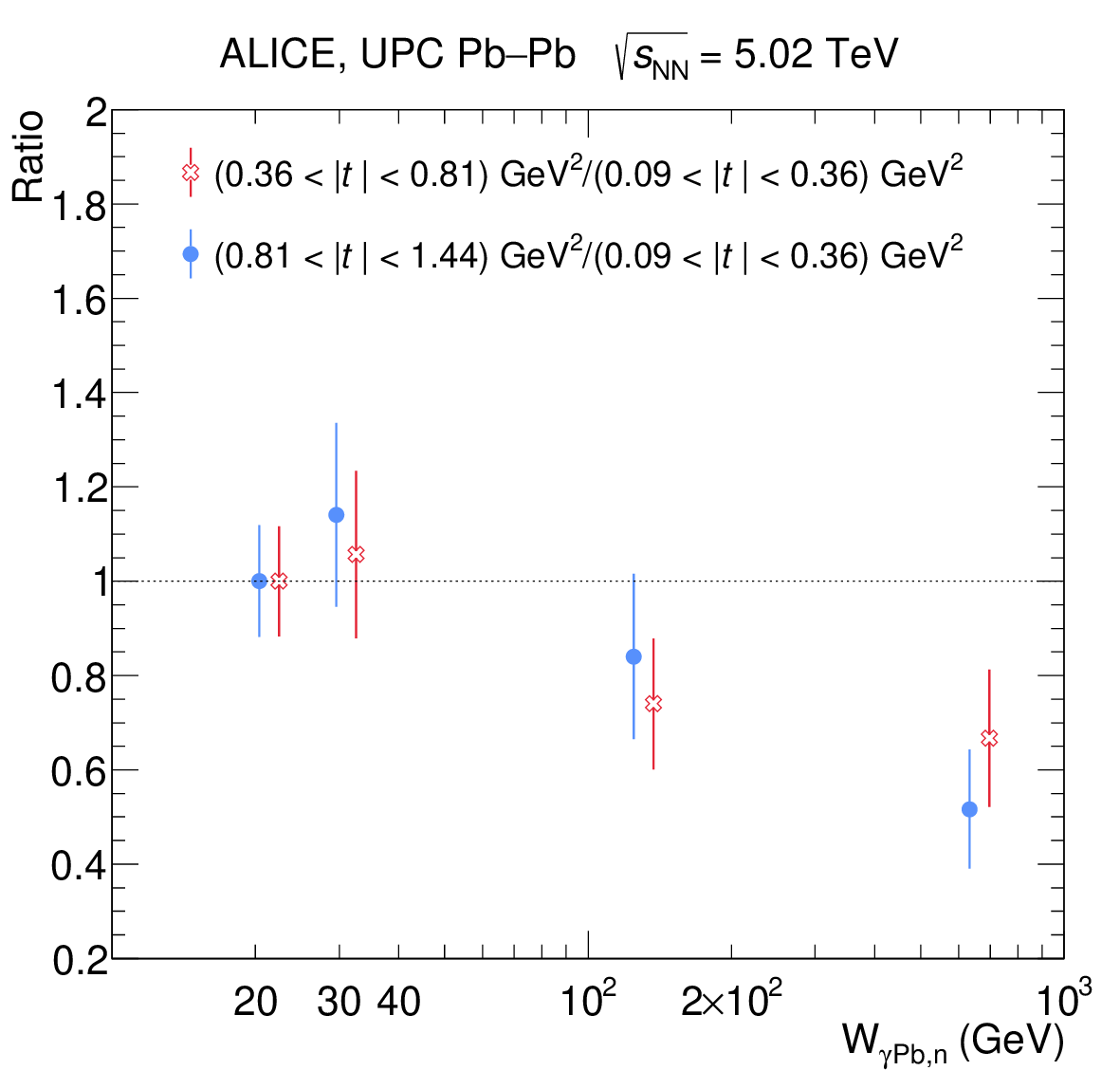}
\caption{The energy dependence of the cross section of incoherent \Jpsi photonuclear production off lead nuclei is shown for three different ranges of the Mandelstam-$t$ variable (left). The symbols denote the measured cross section, the vertical lines across them represent the uncorrelated uncertainty, the open boxes the correlated uncertainty, and the horizontal shadow box the uncertainty from the computation of the photon flux. Ratio of the cross sections at different Mandelstam-$t$ ranges (right) normalised such that the lower energy ratio is one. The vertical line denotes the statistical and \mant-uncorrelated systematic uncertainties added in
quadrature. In the left panel the data symbols have been slightly displaced horizontally to improve their visibility.
}
\label{fig:xs}
\end{figure}

The cross section for incoherent photonuclear production of \Jpsi as a function of $\WgPb$, shown in Fig.~\ref{fig:xs}, displays an interesting trend over the three \mant intervals probed. For the lower \mant values, there is a visible increase in the cross section as the energy increases. A similar, but softer, trend is also observed within the intermediate \mant values. Regarding the data for the uppermost \mant values, the cross section initially rises with energy, reaching a maximum, and then stops growing at the largest \WgPb.
As mentioned above, \mant is related to spatial size scales within the impact-parameter plane. Specifically, a value of $\mant=0.1$ \GeVsqr corresponds to scales of approximately 0.6 fm in size, while $\mant=1.0$ \GeVsqr is dominated by the contribution of structures with a scale of the order of 0.2 fm. To effectively quantify the energy dependence of the cross section across these different scales, the ratio of cross sections with different \mant ranges, normalised such that this ratio is unity at $\WgPb=20$~\GeV, is studied. This ratio is shown in Fig.~\ref{fig:xs}, right. At the highest energies and \mant values the ratio is $0.52\pm0.13$, well above 3 standard deviations from unity, where statistical and uncorrelated systematic uncertainties from the normalisation factor at low energies and the measurements at high energies were added in quadrature. This quantifies the suppression of the growth of incoherent \Jpsi production with energy when sampling small size
gluon configurations in the transverse plane, with respect to sampling larger size regions.

\begin{figure}[!t]
\centering
\includegraphics[width=0.98\textwidth]{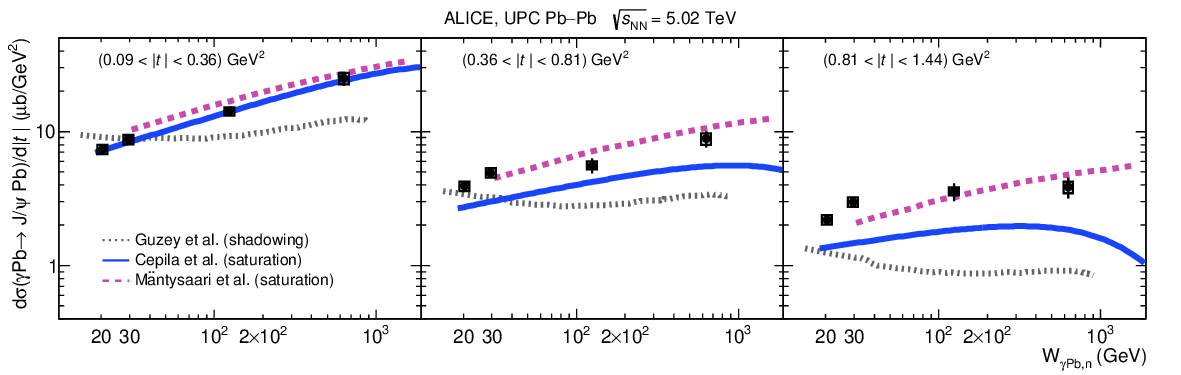}
\caption{The energy dependence of the cross section of incoherent \Jpsi photonuclear production off lead nuclei is shown for three ranges of the Mandelstam-$t$ variable and compared with model predictions. The shadowing based model of Guzey et al.~\cite{PhysRevC.99.015201} is shown with a dot-dashed line. Two saturation-based predictions are also shown: from M{\"a}ntysaari et al.~\cite{Mantysaari:2022sux,Mantysaari:2023xcu} with a dashed line and from Cepila et al.~\cite{Cepila:2023dxn} with a solid line. 
}
\label{fig:models}
\end{figure}

In order to explore the origin of this suppression, data are compared with model predictions in Fig.~\ref{fig:models}. In the leading twist approximation (LTA) by Guzey et al., shadowing is described by multiple scattering of the probe amongst different nucleons in the target~\cite{Frankfurt:2011cs}. This model offers two extreme cases, given by the uncertainty on the determination of the parameter that models three or more nucleon interactions, representing the cases of weak and strong shadowing. When compared with the \WgPb dependence of coherent production of \Jpsi, the average of these two cases successfully describes the \mant-integrated measurements above $\WgPb \sim 100$~GeV~\cite{ALICE:2023jgu,CMS:2023snh}. At $\WgPb \approx 500$~GeV, their predictions differ by about a factor of two for coherent production and three for incoherent production. With a $\sim$20\% experimental uncertainty on the coherent cross section in this range, the measurement strongly constrains the model, and the average of the two scenarios is taken as the LTA reference. 
LTA can also predict the behaviour of incoherent photonuclear production of \Jpsi at a fixed energy of $\WgPb=125$~GeV~\cite{PhysRevC.99.015201}. In this case, the cross section is given by the product of a shadowing factor, that does not depend on \mant, and a parameterisation of HERA data on exclusive and dissociative \Jpsi photoproduction. At energies exceeding 100~GeV, the difference between the model and the measurements is above 3 sigmas (adding all uncertainties in quadrature) for the lower \mant ranges and above 4 sigmas for the large \mant range. Thus, this implementation of nuclear shadowing, which otherwise successfully describes the coherent process, is disfavoured by the data.

The data are also compared with two saturation models, which are based on similar underlying physics concepts but incorporate differing levels of fluctuations in the colour field of Pb. First, data are compared with the energy-dependent hot-spot model by Cepila et al.~\cite{Cepila:2023dxn}, where the energy evolution is given by two sources: the Golec-Biernat W\"usthoff model~\cite{Golec-Biernat:1998zce} and the growth in the number of hot spots inside a nucleon to mimic the growth of the gluon density with energy~\cite{Cepila:2016uku}. The position of the hot spots within the nucleon changes event by event. The model by Cepila et al. is framed within the Good-Walker (GW) approach~\cite{Good:1960ba,Miettinen:1978jb,Klein:2023zlf} where the coherent process is proportional to the average of colour configurations in the target, while the incoherent production is sensitive to the variance of those configurations. 
In this model the decrease of the cross section seen at large energies is a geometric
effect~\cite{Bendova:2018bbb}, given by the overlap of hot spots in the transverse plane when their number is large.  Thus, as the incoherent cross section is given by the variance over colour configurations of the target, the observed decrease implies that all configurations at the corresponding size scale start to resemble each other, as expected when the saturation regime is approached.
As illustrated in Fig.~\ref{fig:models}, this model predicts a power-law growth with energy at small \mant. However, at large \mant, the power-law behaviour is evident only at lower energies, while the cross section is predicted to {\em decrease} at higher energies. For energies above 100~GeV, the difference between the theoretical prediction and the data is below 2 sigmas for the two lower \mant ranges and around 2.5 sigmas for the large \mant range.

Another saturation based model formulated within the GW approach is that by M{\"a}ntysaari et al.~\cite{Mantysaari:2022sux,Mantysaari:2023xcu}. They implement the energy evolution by solving the JIMWLK equation~\cite{Mueller:2001uk}, where the initial condition incorporates nucleon-shape fluctuations using three hot spots whose position and colour strength varies event-by-event~\cite{Mantysaari:2016ykx}. 
Figure~\ref{fig:models} demonstrates that this model predicts a power-law behaviour across the three \mant ranges, with the power-law exponent decreasing as \mant increases. Beyond the energies of 100~GeV, the variation between the model and the data is below or around 2 sigmas for all data.  First studies on the uncertainties in dipole models indicate that they are at the \(5\text{--}10\%\) level, with no strong dependence on kinematics; see e.g.\ Fig.~4\,(\text{b},\text{d}), orange line in Ref.~\cite{Mantysaari:2025ltq}. The results from Refs.~\cite{Casuga:2025etc,Gao:2025dkn}, although comparing only to DIS, yield even smaller uncertainties.

The shadowing and saturation models yield distinct predictions for the first time, despite both successfully describing the cross section for coherent \Jpsi photonuclear production at high energies~\cite{ALICE:2023jgu}. The current LTA implementation of the shadowing model is disfavoured by this analysis, highlighting the need for comprehensive comparisons between theoretical predictions and data, including both coherent and incoherent \Jpsi production.

Regarding the saturation models discussed here, as anticipated, they exhibit a distinct behaviour at high energies and large \mant values. Additional data in this kinematic region would be invaluable for elucidating the underlying physical mechanisms driving the energy evolution of the cross section at large \mant. Recent work by M\"antysaari et al. has suggested that their saturation model requires the inclusion of enhanced quantum fluctuation sources~\cite{Mantysaari:2022sux}. These experimental results call for new theoretical studies on how quantum fluctuations are incorporated in saturation  models. 

In summary, for the first time, the incoherent \Jpsi photonuclear cross section was measured as a function of both \WgPb and \mant. The measurement extends from $\WgPb=20$~GeV to $\WgPb=633$~GeV, and it was carried out in three different \mant ranges. The rise of the cross section with energy shows evidence of suppression as \mant increases. For $\WgPb = 633$~GeV, the ratio between cross sections at the lowest and the highest \mant interval is $0.52\pm 0.13$, a three standard deviation effect. The rise of the cross section with energy exhibits clear suppression at large \mant with respect to low \mant. 
The energy dependence of this suppression, compared with the models discussed here, suggests that gluon saturation plays a significant role at smaller spatial scales.


\newenvironment{acknowledgement}{\relax}{\relax}
\begin{acknowledgement}
\section*{Acknowledgements}

The ALICE Collaboration would like to thank all its engineers and technicians for their invaluable contributions to the construction of the experiment and the CERN accelerator teams for the outstanding performance of the LHC complex.
The ALICE Collaboration gratefully acknowledges the resources and support provided by all Grid centres and the Worldwide LHC Computing Grid (WLCG) collaboration.
The ALICE Collaboration acknowledges the following funding agencies for their support in building and running the ALICE detector:
A. I. Alikhanyan National Science Laboratory (Yerevan Physics Institute) Foundation (ANSL), State Committee of Science and World Federation of Scientists (WFS), Armenia;
Austrian Academy of Sciences, Austrian Science Fund (FWF): [M 2467-N36] and Nationalstiftung f\"{u}r Forschung, Technologie und Entwicklung, Austria;
Ministry of Communications and High Technologies, National Nuclear Research Center, Azerbaijan;
Conselho Nacional de Desenvolvimento Cient\'{\i}fico e Tecnol\'{o}gico (CNPq), Financiadora de Estudos e Projetos (Finep), Funda\c{c}\~{a}o de Amparo \`{a} Pesquisa do Estado de S\~{a}o Paulo (FAPESP) and Universidade Federal do Rio Grande do Sul (UFRGS), Brazil;
Bulgarian Ministry of Education and Science, within the National Roadmap for Research Infrastructures 2020-2027 (object CERN), Bulgaria;
Ministry of Education of China (MOEC) , Ministry of Science \& Technology of China (MSTC) and National Natural Science Foundation of China (NSFC), China;
Ministry of Science and Education and Croatian Science Foundation, Croatia;
Centro de Aplicaciones Tecnol\'{o}gicas y Desarrollo Nuclear (CEADEN), Cubaenerg\'{\i}a, Cuba;
Ministry of Education, Youth and Sports of the Czech Republic, Czech Republic;
The Danish Council for Independent Research | Natural Sciences, the VILLUM FONDEN and Danish National Research Foundation (DNRF), Denmark;
Helsinki Institute of Physics (HIP), Finland;
Commissariat \`{a} l'Energie Atomique (CEA) and Institut National de Physique Nucl\'{e}aire et de Physique des Particules (IN2P3) and Centre National de la Recherche Scientifique (CNRS), France;
Bundesministerium f\"ur Forschung, Technologie und Raumfahrt (BMFTR)and GSI Helmholtzzentrum f\"{u}r Schwerionenforschung GmbH, Germany;
General Secretariat for Research and Technology, Ministry of Education, Research and Religions, Greece;
National Research, Development and Innovation Office, Hungary;
Department of Atomic Energy Government of India (DAE), Department of Science and Technology, Government of India (DST), University Grants Commission, Government of India (UGC) and Council of Scientific and Industrial Research (CSIR), India;
National Research and Innovation Agency - BRIN, Indonesia;
Istituto Nazionale di Fisica Nucleare (INFN), Italy;
Japanese Ministry of Education, Culture, Sports, Science and Technology (MEXT) and Japan Society for the Promotion of Science (JSPS) KAKENHI, Japan;
Consejo Nacional de Ciencia (CONACYT) y Tecnolog\'{i}a, through Fondo de Cooperaci\'{o}n Internacional en Ciencia y Tecnolog\'{i}a (FONCICYT) and Direcci\'{o}n General de Asuntos del Personal Academico (DGAPA), Mexico;
Nederlandse Organisatie voor Wetenschappelijk Onderzoek (NWO), Netherlands;
The Research Council of Norway, Norway;
Pontificia Universidad Cat\'{o}lica del Per\'{u}, Peru;
Ministry of Science and Higher Education, National Science Centre and WUT ID-UB, Poland;
Korea Institute of Science and Technology Information and National Research Foundation of Korea (NRF), Republic of Korea;
Ministry of Education and Scientific Research, Institute of Atomic Physics, Ministry of Research and Innovation and Institute of Atomic Physics and Universitatea Nationala de Stiinta si Tehnologie Politehnica Bucuresti, Romania;
Ministerstvo skolstva, vyskumu, vyvoja a mladeze SR, Slovakia;
National Research Foundation of South Africa, South Africa;
Swedish Research Council (VR) and Knut \& Alice Wallenberg Foundation (KAW), Sweden;
European Organization for Nuclear Research, Switzerland;
Suranaree University of Technology (SUT), National Science and Technology Development Agency (NSTDA) and National Science, Research and Innovation Fund (NSRF via PMU-B B05F650021), Thailand;
Turkish Energy, Nuclear and Mineral Research Agency (TENMAK), Turkey;
National Academy of  Sciences of Ukraine, Ukraine;
Science and Technology Facilities Council (STFC), United Kingdom;
National Science Foundation of the United States of America (NSF) and United States Department of Energy, Office of Nuclear Physics (DOE NP), United States of America.
In addition, individual groups or members have received support from:
Czech Science Foundation (grant no. 23-07499S), Czech Republic;
FORTE project, reg.\ no.\ CZ.02.01.01/00/22\_008/0004632, Czech Republic, co-funded by the European Union, Czech Republic;
European Research Council (grant no. 950692), European Union;
Academy of Finland (Center of Excellence in Quark Matter) (grant nos. 346327, 346328), Finland;
Deutsche Forschungs Gemeinschaft (DFG, German Research Foundation) ``Neutrinos and Dark Matter in Astro- and Particle Physics'' (grant no. SFB 1258), Germany;
ICSC - National Research Center for High Performance Computing, Big Data and Quantum Computing and FAIR - Future Artificial Intelligence Research, funded by the NextGenerationEU program (Italy).

\end{acknowledgement}

\bibliographystyle{utphys}  
\bibliography{bibliography}

\newpage
\appendix

\section{End Matter}

\subsection{Invariant mass distributions}
\begin{figure}[!htbp]
\centering
\includegraphics[width=0.32\textwidth]{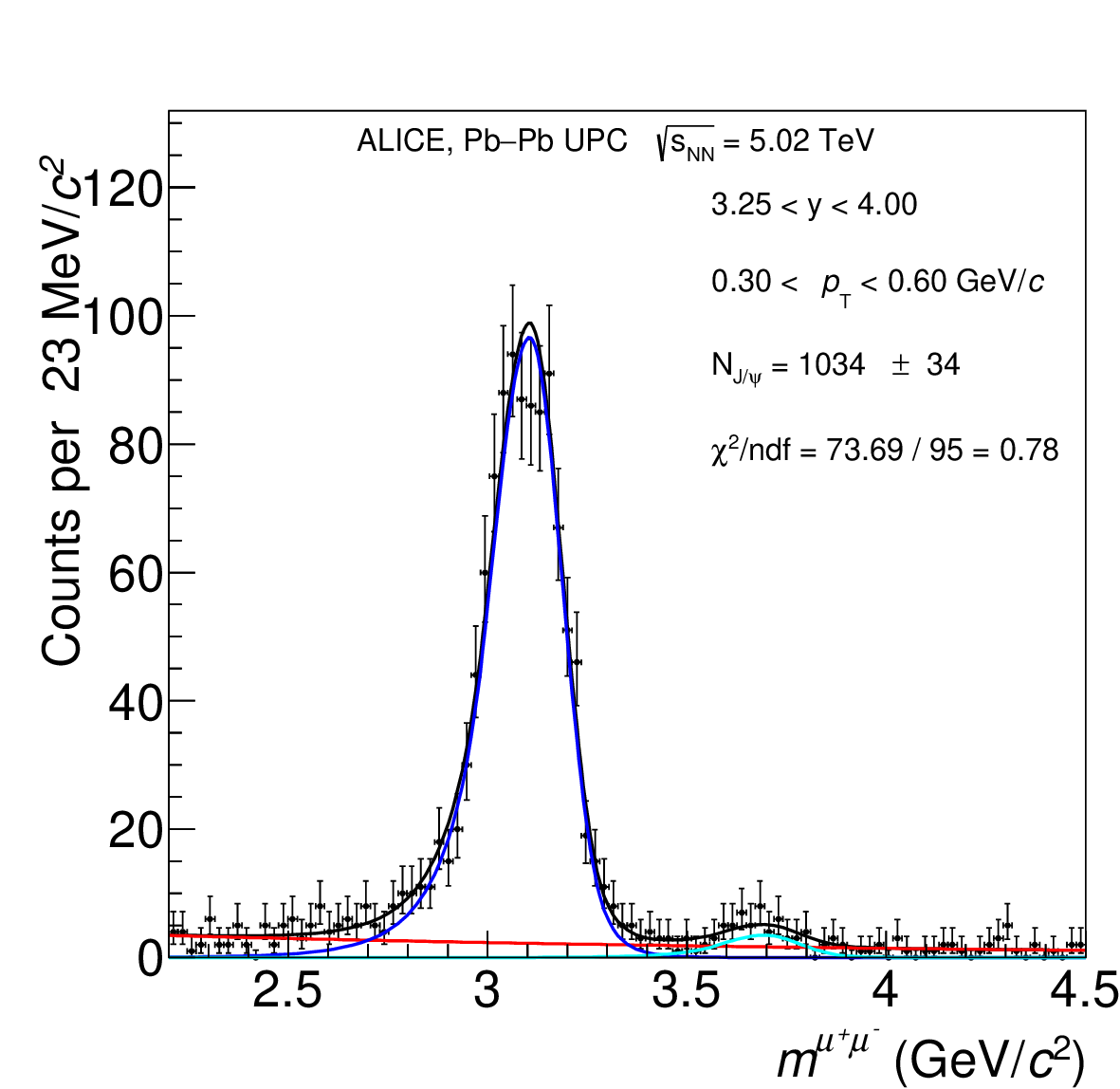}
\includegraphics[width=0.32\textwidth]{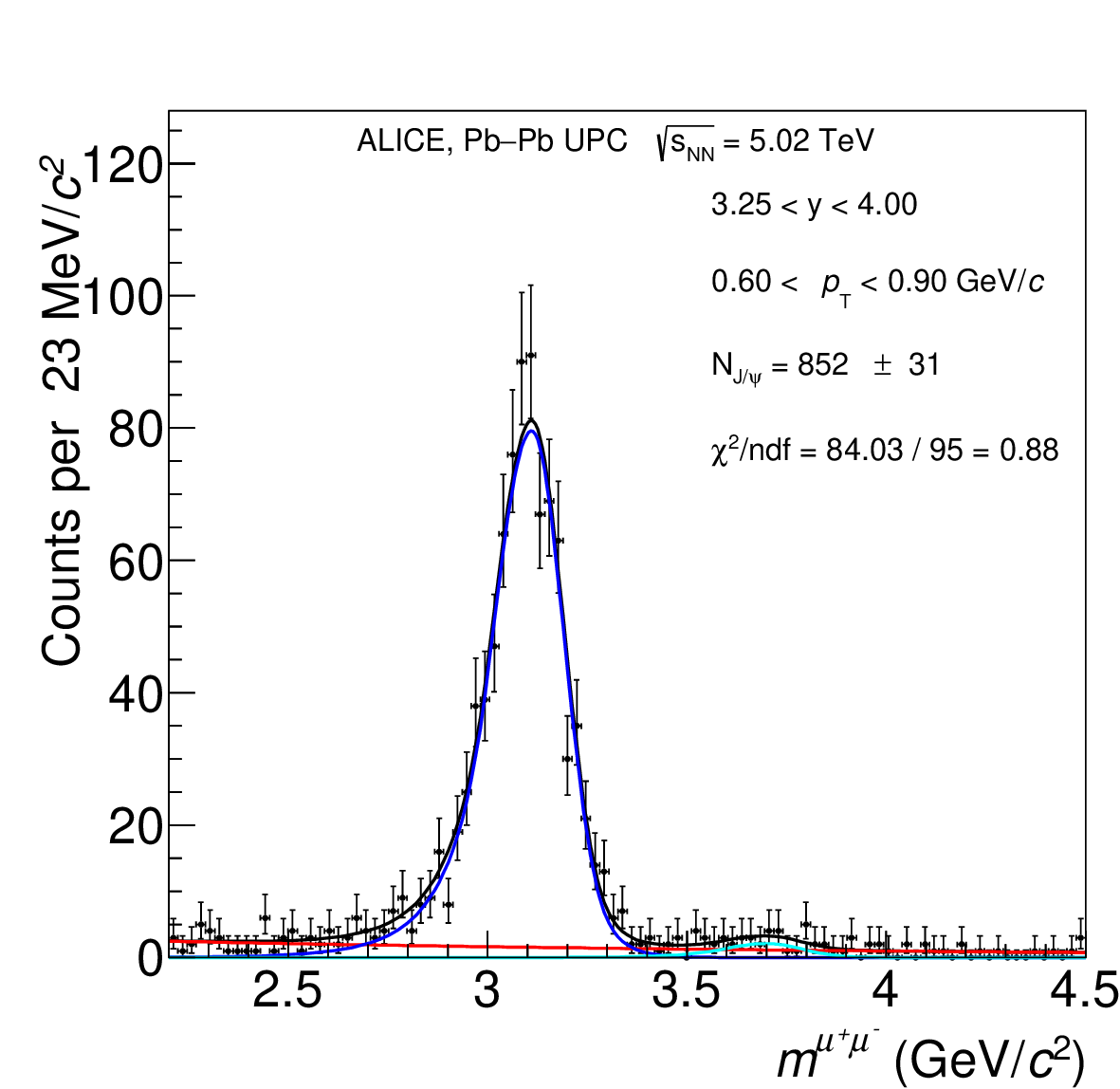}
\includegraphics[width=0.32\textwidth]{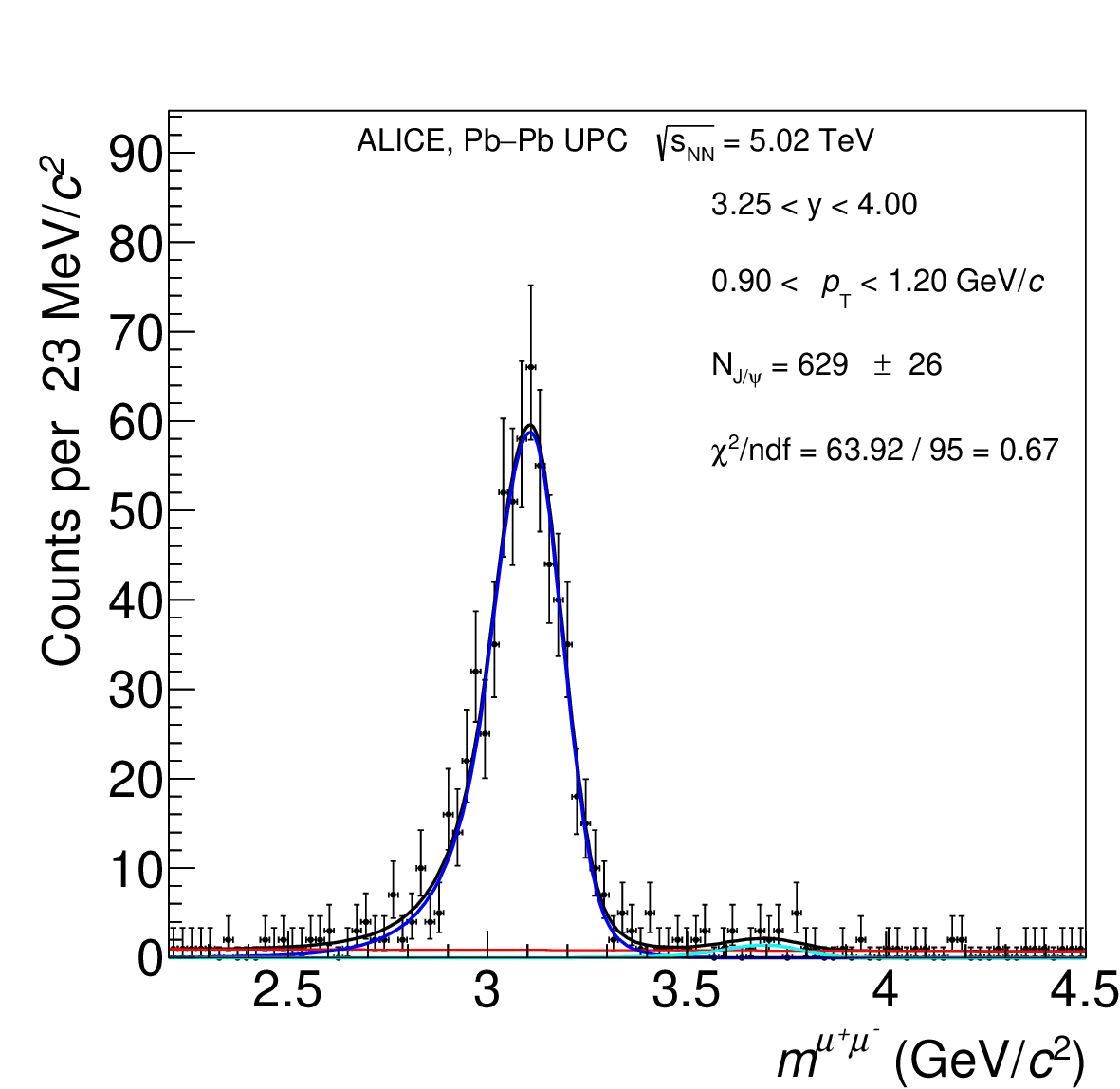}\\
\includegraphics[width=0.32\textwidth]{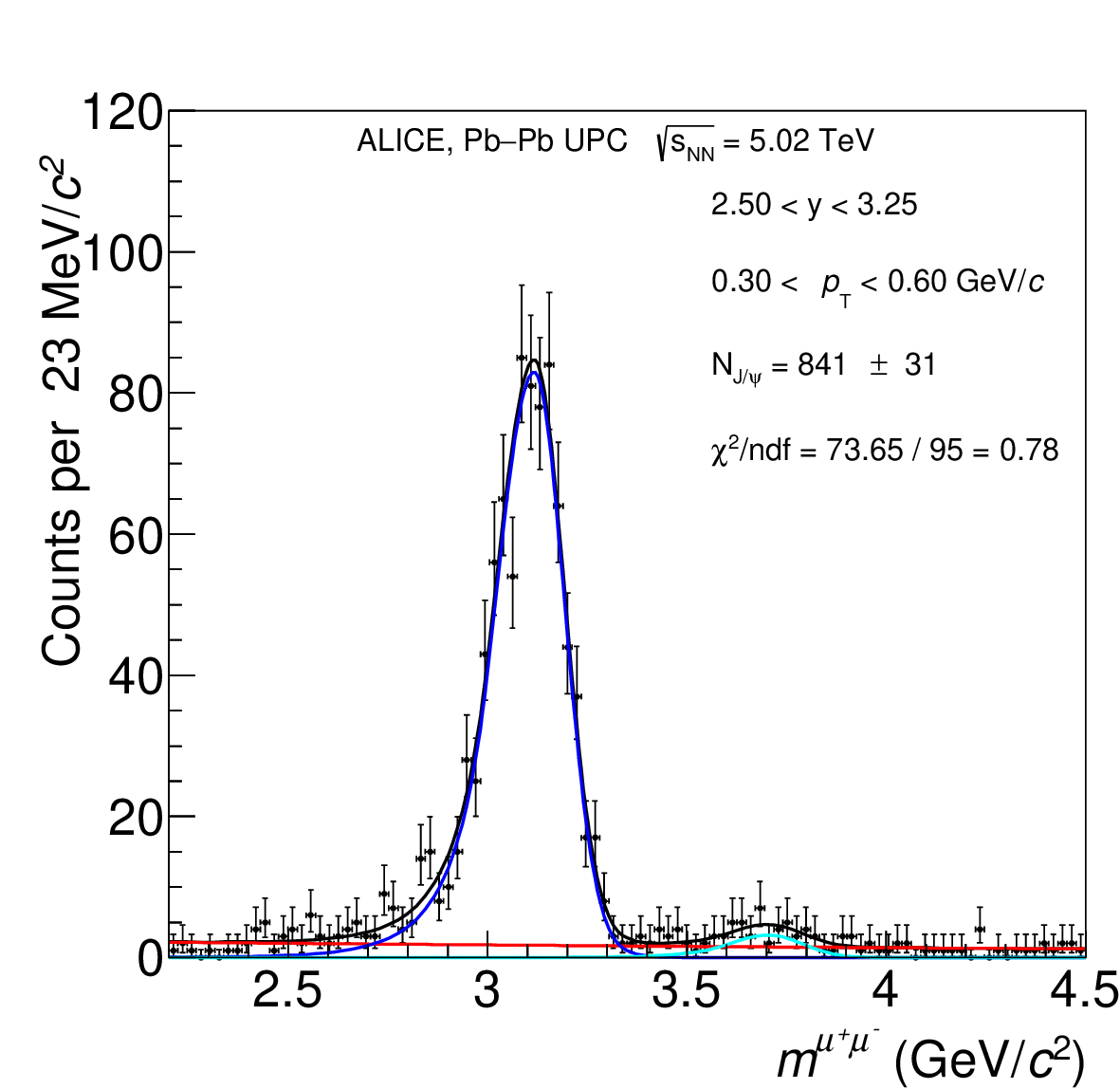}
\includegraphics[width=0.32\textwidth]{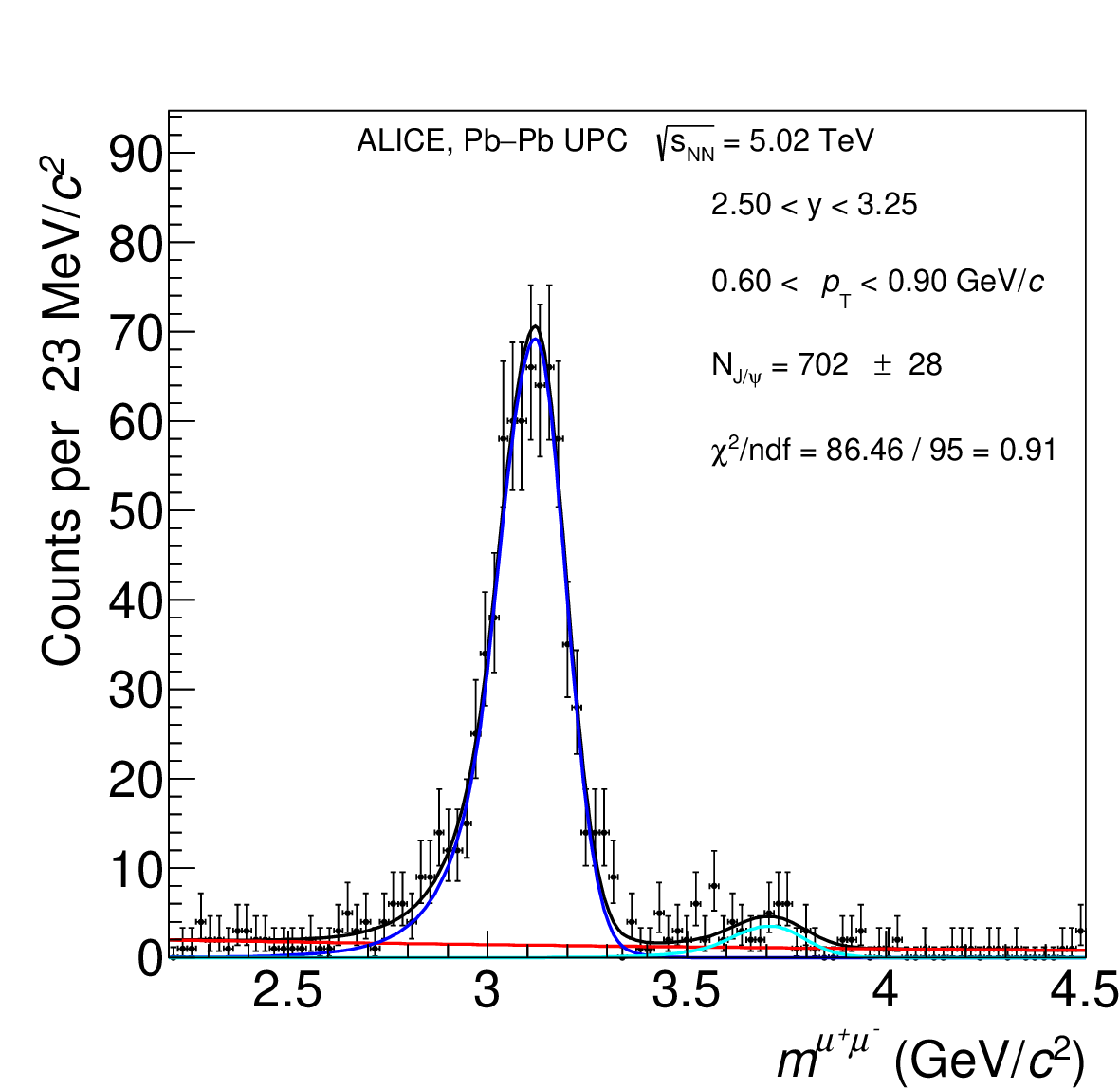}
\includegraphics[width=0.32\textwidth]{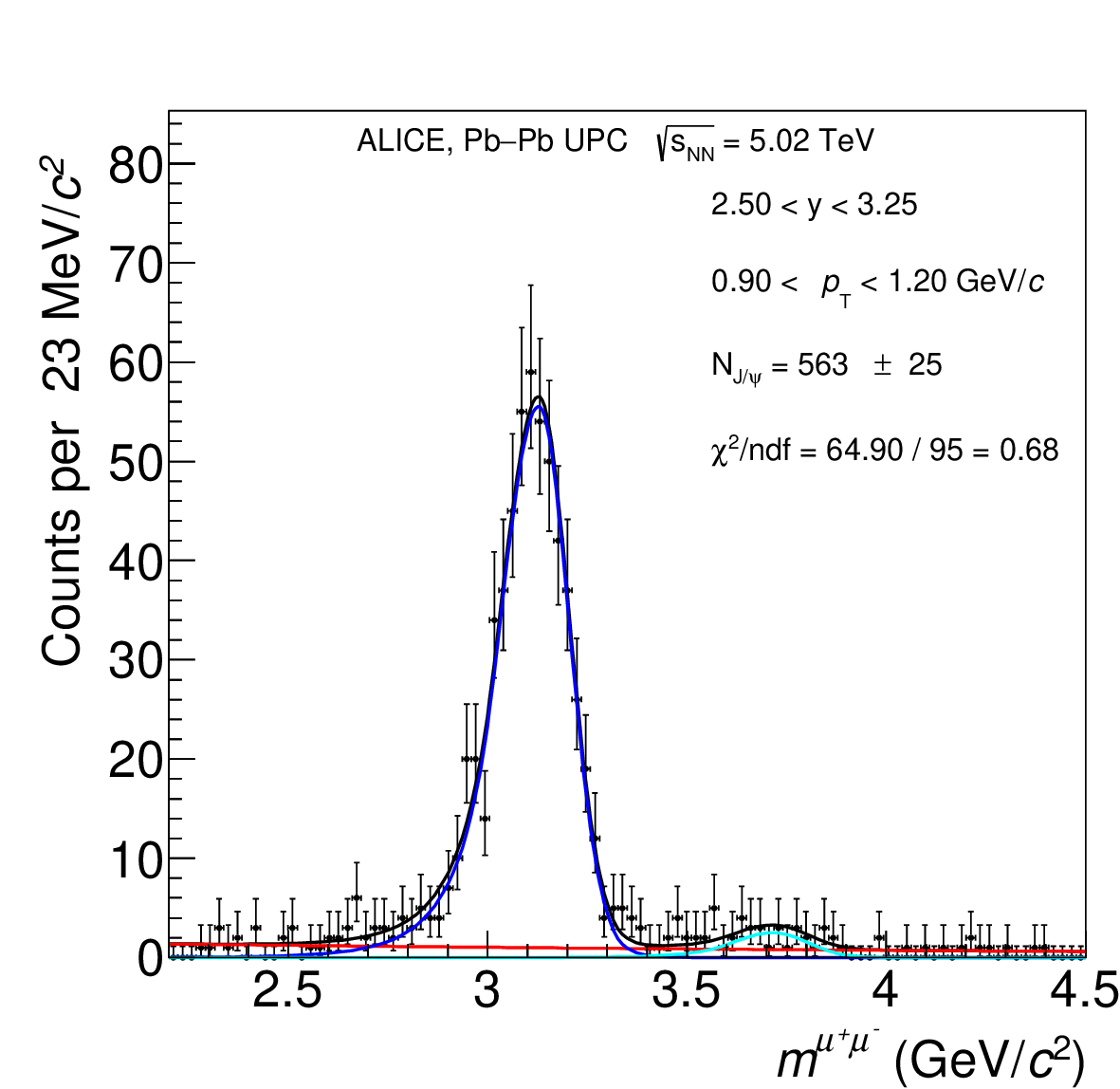}\\
\includegraphics[width=0.32\textwidth]{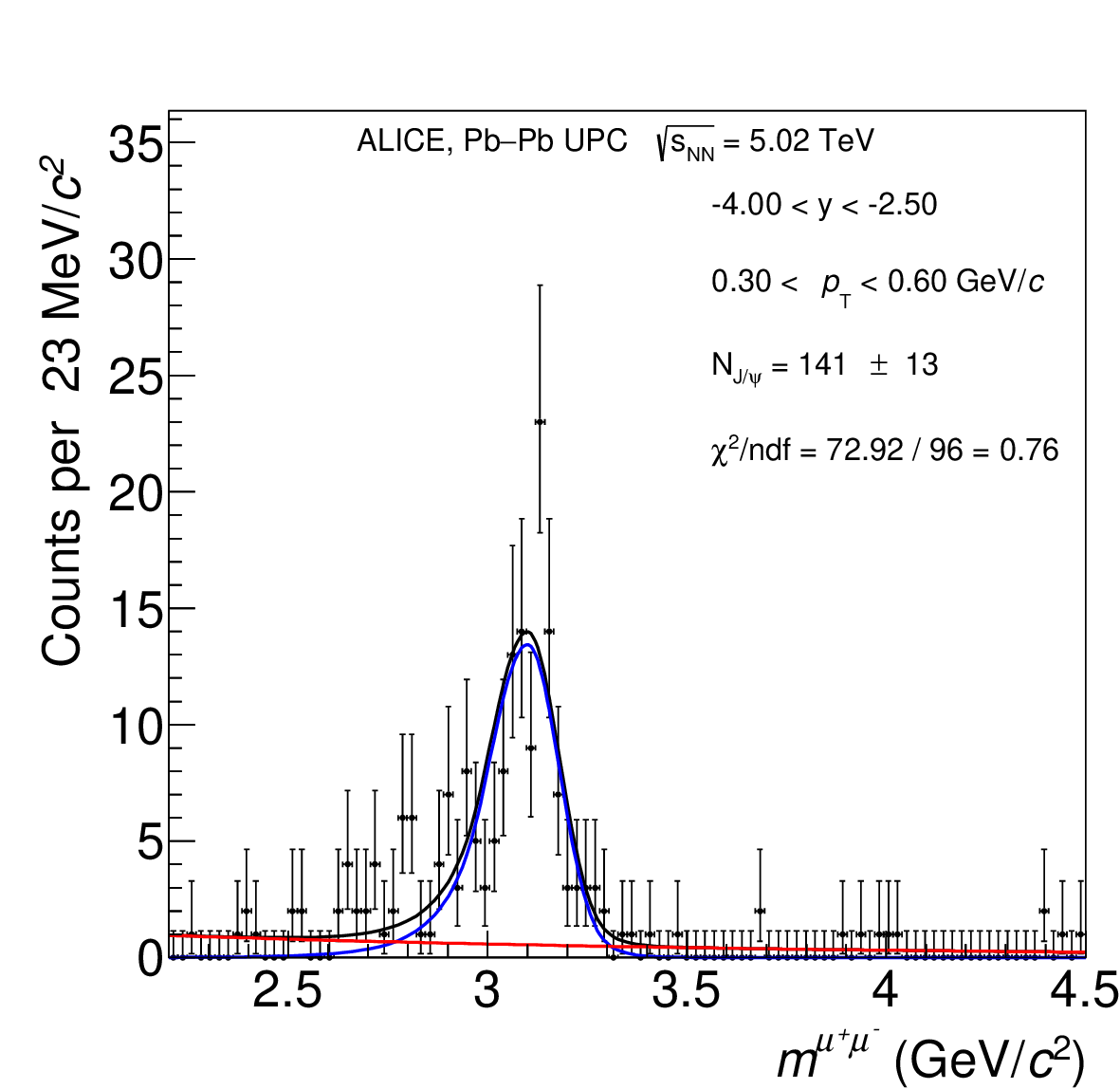}
\includegraphics[width=0.32\textwidth]{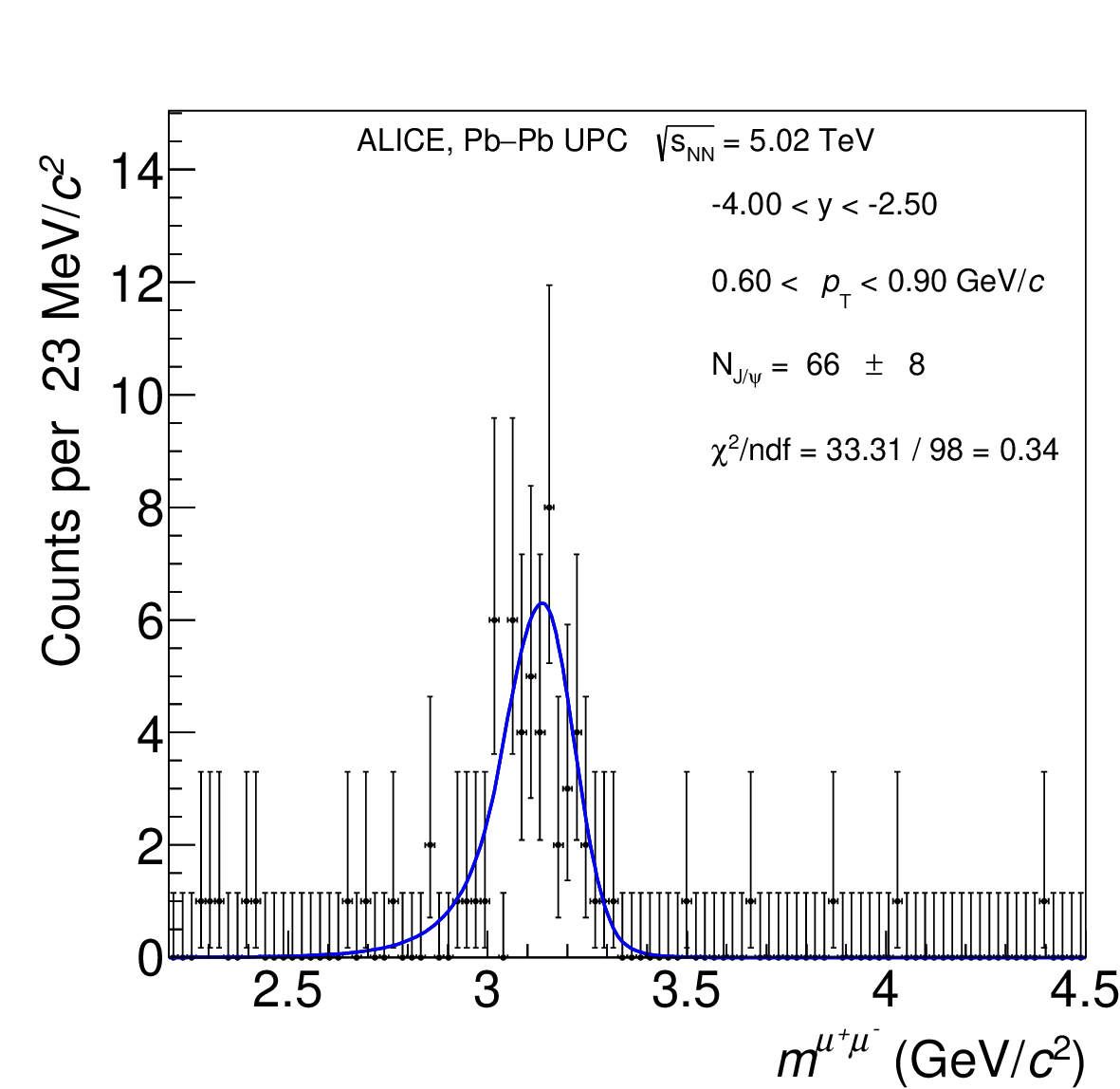}
\includegraphics[width=0.32\textwidth]{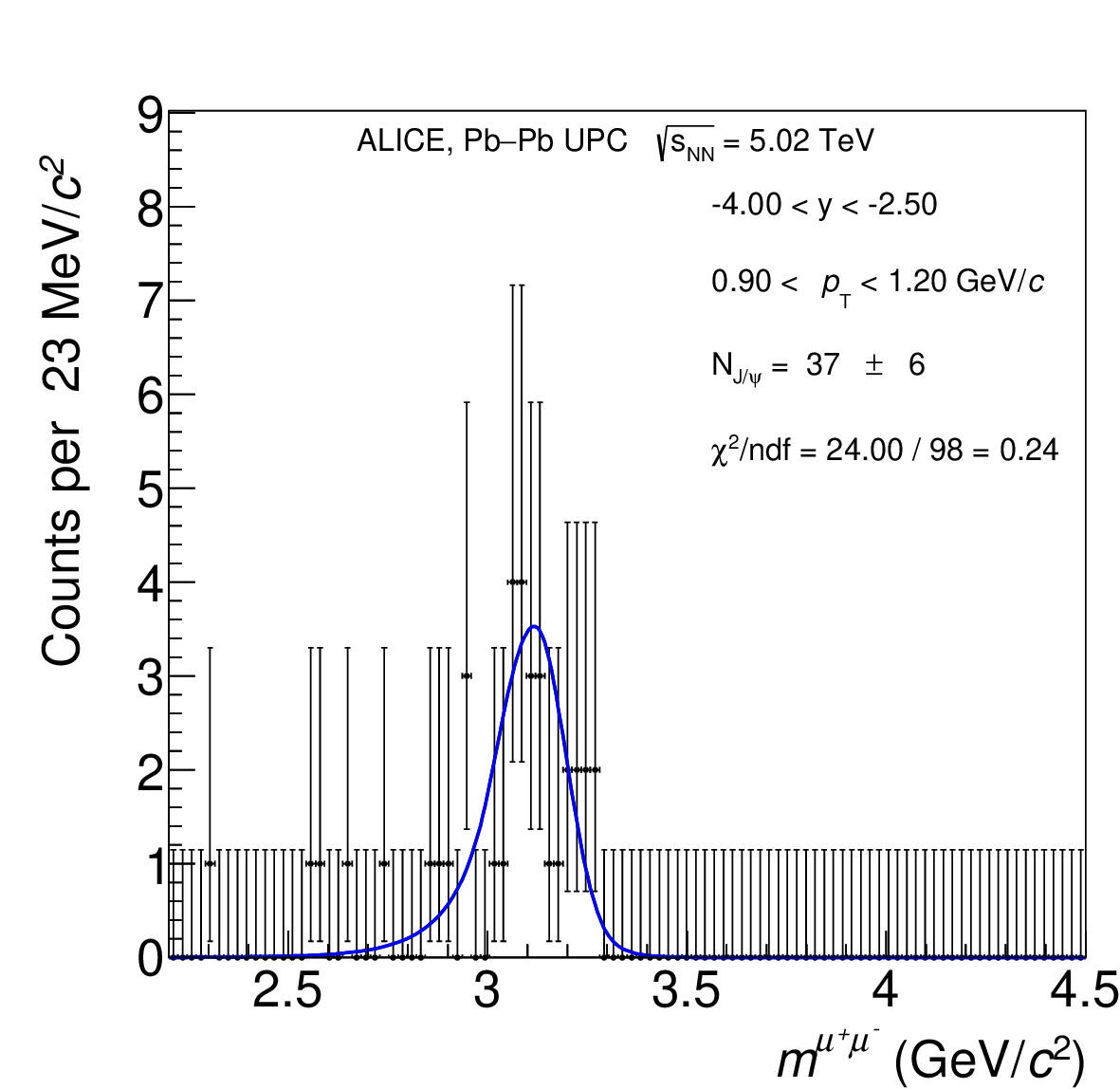}
\caption{Dimuon invariant mass distributions in \pt and y intervals and the corresponding extended unbinned likelihood  fits. The value of $\chi^2$, which is not defined for unbinned fits, is provided for illustration purposes only and correspond to the chosen binning scheme of this figure. 
}
\label{fig:mass}
\end{figure}

The fits to the invariant mass distributions are shown in Fig.~\ref{fig:mass}. The upper set of panels corresponds to the rapidity interval $3.25<y<4.0$, which covers the lowest \WgPb within the measured transverse momentum ranges. In this region, the \Jpsi signal was modelled with a double-sided Crystal Ball function (DSCB), while the \psip contribution and the background were described with an additional DSCB and an exponential function, respectively. The middle row shows the results for $2.5<y<3.25$, where the same fitting procedure was applied. The bottom row corresponds to the backward rapidity interval $-4<y<-2.5$; here, no \psip contribution was included in the fit, and for the last two \pt ranges the background component was not included, as no significant background is present.
\subsection{Transverse momentum distributions}
\begin{figure}[!htbp]
\centering
\includegraphics[width=0.32\textwidth]{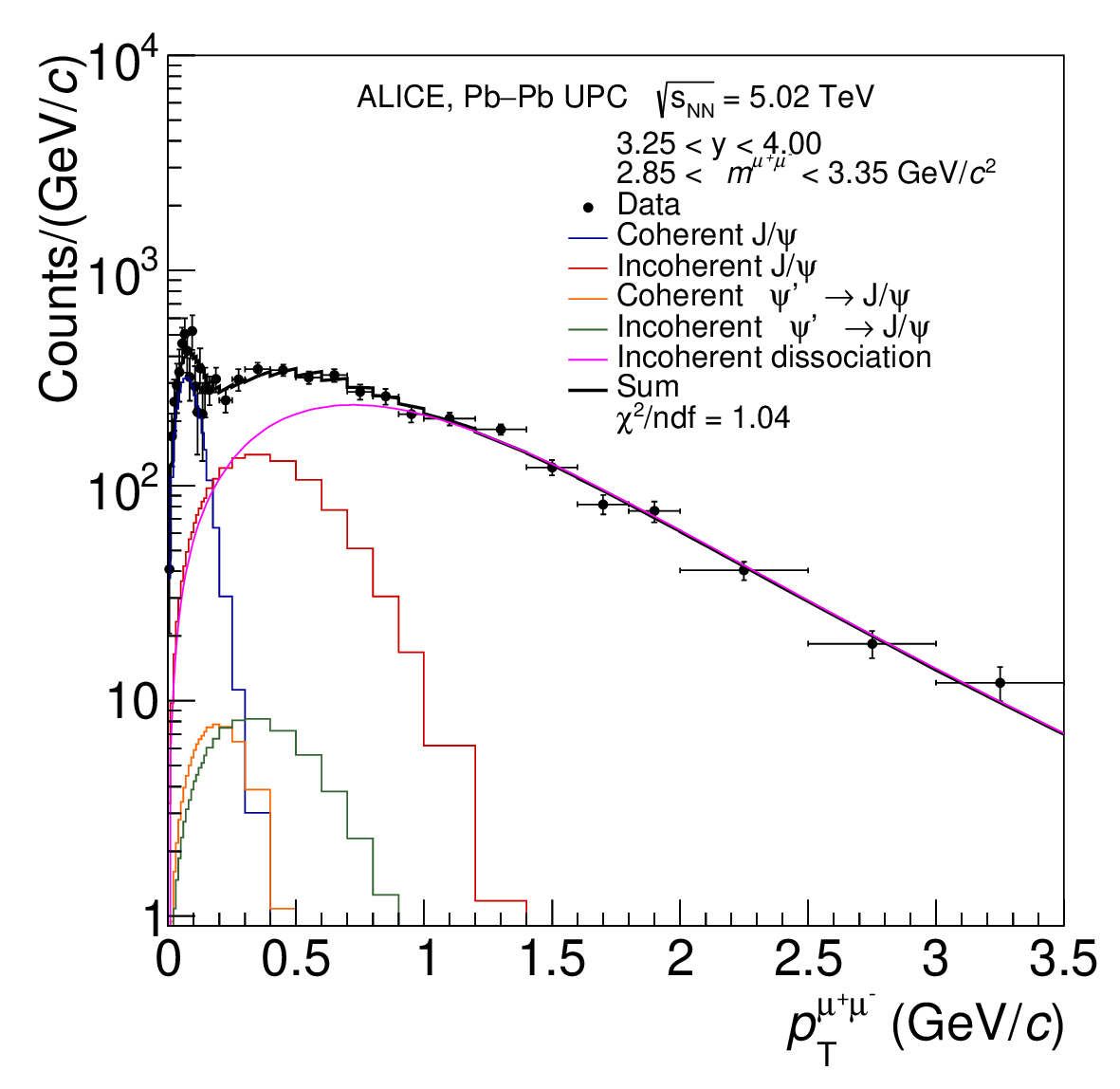}
\includegraphics[width=0.32\textwidth]{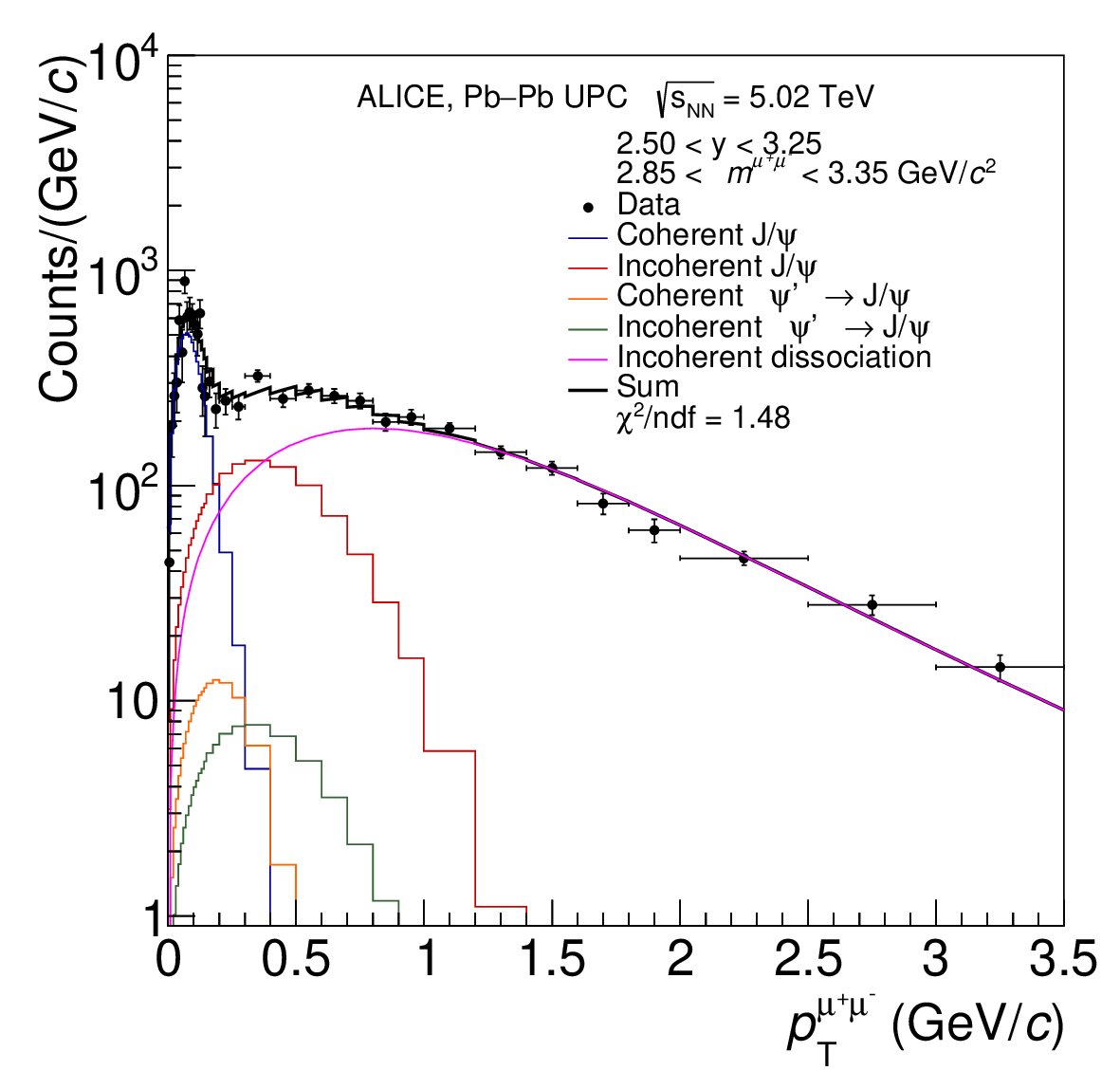}
\includegraphics[width=0.32\textwidth]{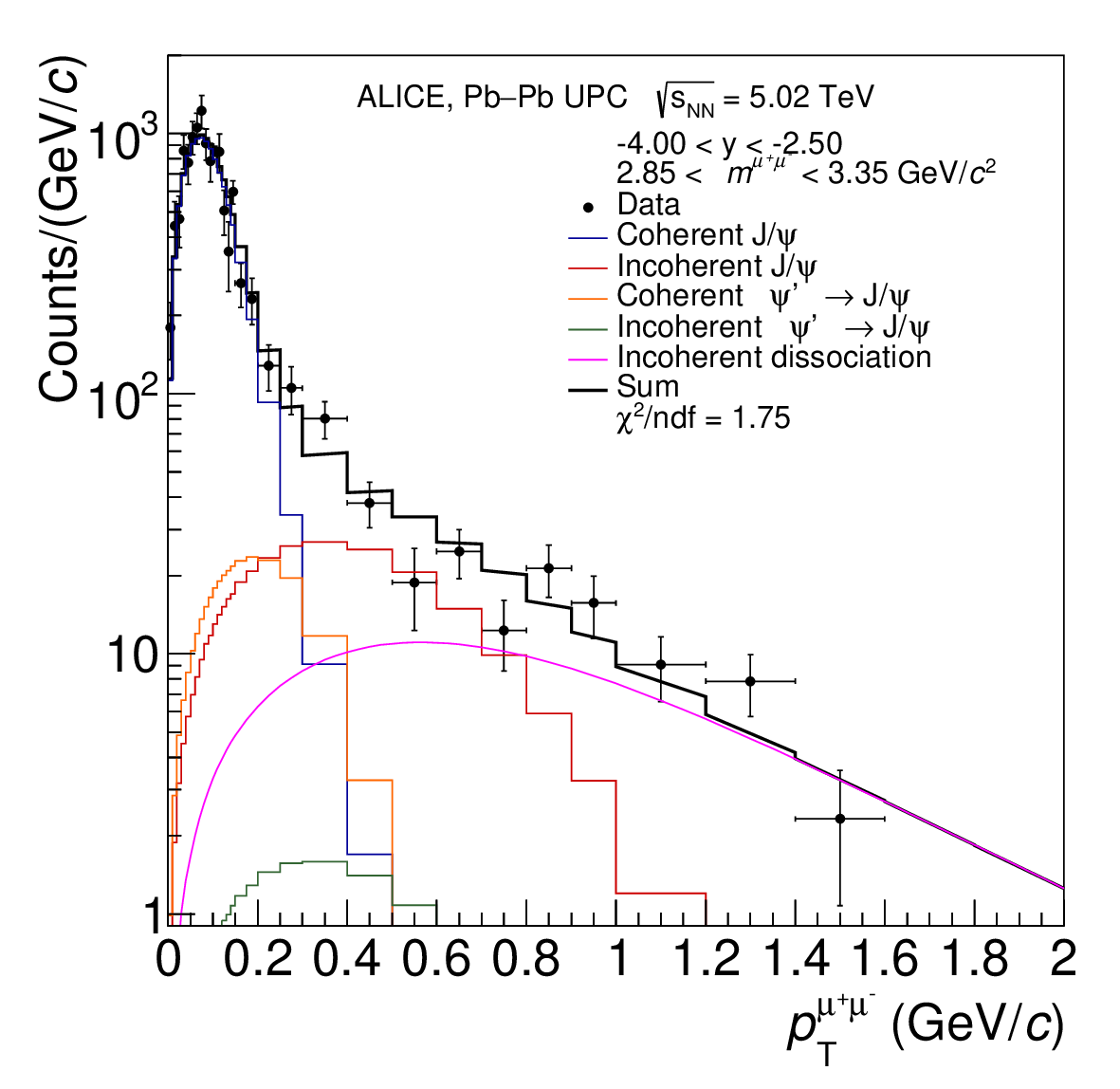}\\
\caption{Transverse momentum distribution of the dimuons for the three rapidity intervals studied with the muon spectrometer.
}
\label{fig:pt}
\end{figure}

Figure~\ref{fig:pt} shows the transverse momentum distribution of the dimuons for the three rapidity intervals studied with the muon spectrometer. It also shows the fit to the different templates that are used to describe the data: contributions from coherent and incoherent elastic production of \Jpsi, the contributions (via feed-down) of the same processes for the case of \psip, and the contribution from incoherent dissociative \Jpsi production. The first 4 templates are obtained from STARlight, while the last template is defined according to the H1 prescription~\cite{H1:2013okq}.
\subsection{Treatment of neutron emission and photon fluxes}

In ultra-peripheral Pb--Pb collisions, neutrons detected in the Zero Degree Calorimeters (ZDCs) originate from two distinct mechanisms:  
(i) \textbf{Electromagnetic Dissociation (EMD)}, where additional photons exchanged between the nuclei lead to nuclear breakup, and  
(ii) \textbf{the hadronic vertex of incoherent $J/\psi$ production}, which, in the $|t|$ range considered here, always produces at least one forward neutron.  

The ALICE ZDC system categorizes events according to neutron detection on each side of the interaction point: \textbf{0n0n}, \textbf{0nXn}, \textbf{Xn0n}, and \textbf{XnXn}, where ``n'' denotes no detected neutron and ``Xn'' indicates one or more detected neutrons.  
These four categories are crucial for separating coherent and incoherent production, as well as for isolating contributions from EMD processes.

For both coherent and incoherent $J/\psi$ production, the photon flux used must correspond to the photon participating in the hard scattering.  
Thus, the full photon flux is applied when modeling the production of the $J/\psi$.  
The EMD-specific fluxes are only used when describing the extra photons responsible for neutron emission via electromagnetic dissociation.  

To disentangle these effects and correct for detector inefficiencies, we simultaneously fit the measured yields in all four ZDC categories.  
This data-driven procedure allows us to:
\begin{enumerate}
    \item Correct for losses due to the finite ZDC efficiency,
    \item Account for pile-up from multiple EMD interactions,
    \item Subtract backgrounds from coherent $J/\psi$, coherent $J/\psi$ + EMD, and pure EMD events originating from other ion pairs.
\end{enumerate}

This system of equations does not rely on external models for neutron emission, as all relevant quantities are determined directly from data.  
By construction, the incoherent component is always treated with the full photon flux, while EMD-only processes are modeled with the appropriate EMD fluxes.

\section{The ALICE Collaboration}
\label{app:collab}
\begin{flushleft} 
\small

S.~Acharya\,\orcidlink{0000-0002-9213-5329}\,$^{\rm 50}$, 
A.~Agarwal$^{\rm 133}$, 
G.~Aglieri Rinella\,\orcidlink{0000-0002-9611-3696}\,$^{\rm 32}$, 
L.~Aglietta\,\orcidlink{0009-0003-0763-6802}\,$^{\rm 24}$, 
M.~Agnello\,\orcidlink{0000-0002-0760-5075}\,$^{\rm 29}$, 
N.~Agrawal\,\orcidlink{0000-0003-0348-9836}\,$^{\rm 25}$, 
Z.~Ahammed\,\orcidlink{0000-0001-5241-7412}\,$^{\rm 133}$, 
S.~Ahmad\,\orcidlink{0000-0003-0497-5705}\,$^{\rm 15}$, 
S.U.~Ahn\,\orcidlink{0000-0001-8847-489X}\,$^{\rm 71}$, 
I.~Ahuja\,\orcidlink{0000-0002-4417-1392}\,$^{\rm 36}$
% ...

\section*{Affiliation Notes}

$^{\rm I}$ Deceased\\
$^{\rm II}$ Also at: Max-Planck-Institut fur Physik, Munich, Germany\\
$^{\rm III}$ Also at: Italian National Agency for New Technologies, Energy and Sustainable Economic Development (ENEA), Bologna, Italy\\
$^{\rm IV}$ Also at: Instituto de Fisica da Universidade de Sao Paulo\\
$^{\rm V}$ Also at: Dipartimento DET del Politecnico di Torino, Turin, Italy\\
$^{\rm VI}$ Also at: Department of Applied Physics, Aligarh Muslim University, Aligarh, India\\
$^{\rm VII}$ Also at: Institute of Theoretical Physics, University of Wroclaw, Poland\\
$^{\rm VIII}$ Also at: Facultad de Ciencias, Universidad Nacional Aut\'{o}noma de M\'{e}xico, Mexico City, Mexico\\

\section*{Collaboration Institutes}

$^{1}$ A.I. Alikhanyan National Science Laboratory (Yerevan Physics Institute) Foundation, Yerevan, Armenia\\
$^{2}$ AGH University of Krakow, Cracow, Poland\\
$^{3}$ Bogolyubov Institute for Theoretical Physics, National Academy of Sciences of Ukraine, Kiev, Ukraine\\
$^{4}$ Bose Institute, Department of Physics  and Centre for Astroparticle Physics and Space Science (CAPSS), Kolkata, India\\
$^{5}$ California Polytechnic State University, San Luis Obispo, California, United States\\
$^{6}$ Central China Normal University, Wuhan, China\\
$^{7}$ Centro de Aplicaciones Tecnol\'{o}gicas y Desarrollo Nuclear (CEADEN), Havana, Cuba\\
$^{8}$ Centro de Investigaci\'{o}n y de Estudios Avanzados (CINVESTAV), Mexico City and M\'{e}rida, Mexico\\
$^{9}$ Chicago State University, Chicago, Illinois, United States\\
$^{10}$ China Nuclear Data Center, China Institute of Atomic Energy, Beijing, China\\
$^{11}$ China University of Geosciences, Wuhan, China\\
$^{12}$ Chungbuk National University, Cheongju, Republic of Korea\\
$^{13}$ Comenius University Bratislava, Faculty of Mathematics, Physics and Informatics, Bratislava, Slovak Republic\\
$^{14}$ Creighton University, Omaha, Nebraska, United States\\
$^{15}$ Department of Physics, Aligarh Muslim University, Aligarh, India\\
$^{16}$ Department of Physics, Pusan National University, Pusan, Republic of Korea\\
$^{17}$ Department of Physics, Sejong University, Seoul, Republic of Korea\\
$^{18}$ Department of Physics, University of California, Berkeley, California, United States\\
$^{19}$ Department of Physics, University of Oslo, Oslo, Norway\\
$^{20}$ Department of Physics and Technology, University of Bergen, Bergen, Norway\\
$^{21}$ Dipartimento di Fisica, Universit\`{a} di Pavia, Pavia, Italy\\
$^{22}$ Dipartimento di Fisica dell'Universit\`{a} and Sezione INFN, Cagliari, Italy\\
$^{23}$ Dipartimento di Fisica dell'Universit\`{a} and Sezione INFN, Trieste, Italy\\
$^{24}$ Dipartimento di Fisica dell'Universit\`{a} and Sezione INFN, Turin, Italy\\
$^{25}$ Dipartimento di Fisica e Astronomia dell'Universit\`{a} and Sezione INFN, Bologna, Italy\\
$^{26}$ Dipartimento di Fisica e Astronomia dell'Universit\`{a} and Sezione INFN, Catania, Italy\\
$^{27}$ Dipartimento di Fisica e Astronomia dell'Universit\`{a} and Sezione INFN, Padova, Italy\\
$^{28}$ Dipartimento di Fisica `E.R.~Caianiello' dell'Universit\`{a} and Gruppo Collegato INFN, Salerno, Italy\\
$^{29}$ Dipartimento DISAT del Politecnico and Sezione INFN, Turin, Italy\\
$^{30}$ Dipartimento di Scienze MIFT, Universit\`{a} di Messina, Messina, Italy\\
$^{31}$ Dipartimento Interateneo di Fisica `M.~Merlin' and Sezione INFN, Bari, Italy\\
$^{32}$ European Organization for Nuclear Research (CERN), Geneva, Switzerland\\
$^{33}$ Faculty of Electrical Engineering, Mechanical Engineering and Naval Architecture, University of Split, Split, Croatia\\
$^{34}$ Faculty of Nuclear Sciences and Physical Engineering, Czech Technical University in Prague, Prague, Czech Republic\\
$^{35}$ Faculty of Physics, Sofia University, Sofia, Bulgaria\\
$^{36}$ Faculty of Science, P.J.~\v{S}af\'{a}rik University, Ko\v{s}ice, Slovak Republic\\
$^{37}$ Faculty of Technology, Environmental and Social Sciences, Bergen, Norway\\
$^{38}$ Frankfurt Institute for Advanced Studies, Johann Wolfgang Goethe-Universit\"{a}t Frankfurt, Frankfurt, Germany\\
$^{39}$ Fudan University, Shanghai, China\\
$^{40}$ Gangneung-Wonju National University, Gangneung, Republic of Korea\\
$^{41}$ Gauhati University, Department of Physics, Guwahati, India\\
$^{42}$ Helmholtz-Institut f\"{u}r Strahlen- und Kernphysik, Rheinische Friedrich-Wilhelms-Universit\"{a}t Bonn, Bonn, Germany\\
$^{43}$ Helsinki Institute of Physics (HIP), Helsinki, Finland\\
$^{44}$ High Energy Physics Group,  Universidad Aut\'{o}noma de Puebla, Puebla, Mexico\\
$^{45}$ Horia Hulubei National Institute of Physics and Nuclear Engineering, Bucharest, Romania\\
$^{46}$ HUN-REN Wigner Research Centre for Physics, Budapest, Hungary\\
$^{47}$ Indian Institute of Technology Bombay (IIT), Mumbai, India\\
$^{48}$ Indian Institute of Technology Indore, Indore, India\\
$^{49}$ INFN, Laboratori Nazionali di Frascati, Frascati, Italy\\
$^{50}$ INFN, Sezione di Bari, Bari, Italy\\
$^{51}$ INFN, Sezione di Bologna, Bologna, Italy\\
$^{52}$ INFN, Sezione di Cagliari, Cagliari, Italy\\
$^{53}$ INFN, Sezione di Catania, Catania, Italy\\
$^{54}$ INFN, Sezione di Padova, Padova, Italy\\
$^{55}$ INFN, Sezione di Pavia, Pavia, Italy\\
$^{56}$ INFN, Sezione di Torino, Turin, Italy\\
$^{57}$ INFN, Sezione di Trieste, Trieste, Italy\\
$^{58}$ Inha University, Incheon, Republic of Korea\\
$^{59}$ Institute for Gravitational and Subatomic Physics (GRASP), Utrecht University/Nikhef, Utrecht, Netherlands\\
$^{60}$ Institute of Experimental Physics, Slovak Academy of Sciences, Ko\v{s}ice, Slovak Republic\\
$^{61}$ Institute of Physics, Homi Bhabha National Institute, Bhubaneswar, India\\
$^{62}$ Institute of Physics of the Czech Academy of Sciences, Prague, Czech Republic\\
$^{63}$ Institute of Space Science (ISS), Bucharest, Romania\\
$^{64}$ Institut f\"{u}r Kernphysik, Johann Wolfgang Goethe-Universit\"{a}t Frankfurt, Frankfurt, Germany\\
$^{65}$ Instituto de Ciencias Nucleares, Universidad Nacional Aut\'{o}noma de M\'{e}xico, Mexico City, Mexico\\
$^{66}$ Instituto de F\'{i}sica, Universidade Federal do Rio Grande do Sul (UFRGS), Porto Alegre, Brazil\\
$^{67}$ Instituto de F\'{\i}sica, Universidad Nacional Aut\'{o}noma de M\'{e}xico, Mexico City, Mexico\\
$^{68}$ iThemba LABS, National Research Foundation, Somerset West, South Africa\\
$^{69}$ Jeonbuk National University, Jeonju, Republic of Korea\\
$^{70}$ Johann-Wolfgang-Goethe Universit\"{a}t Frankfurt Institut f\"{u}r Informatik, Fachbereich Informatik und Mathematik, Frankfurt, Germany\\
$^{71}$ Korea Institute of Science and Technology Information, Daejeon, Republic of Korea\\
$^{72}$ Laboratoire de Physique Subatomique et de Cosmologie, Universit\'{e} Grenoble-Alpes, CNRS-IN2P3, Grenoble, France\\
$^{73}$ Lawrence Berkeley National Laboratory, Berkeley, California, United States\\
$^{74}$ Lund University Department of Physics, Division of Particle Physics, Lund, Sweden\\
$^{75}$ Nagasaki Institute of Applied Science, Nagasaki, Japan\\
$^{76}$ Nara Women{'}s University (NWU), Nara, Japan\\
$^{77}$ National and Kapodistrian University of Athens, School of Science, Department of Physics , Athens, Greece\\
$^{78}$ National Centre for Nuclear Research, Warsaw, Poland\\
$^{79}$ National Institute of Science Education and Research, Homi Bhabha National Institute, Jatni, India\\
$^{80}$ National Nuclear Research Center, Baku, Azerbaijan\\
$^{81}$ National Research and Innovation Agency - BRIN, Jakarta, Indonesia\\
$^{82}$ Niels Bohr Institute, University of Copenhagen, Copenhagen, Denmark\\
$^{83}$ Nikhef, National institute for subatomic physics, Amsterdam, Netherlands\\
$^{84}$ Nuclear Physics Group, STFC Daresbury Laboratory, Daresbury, United Kingdom\\
$^{85}$ Nuclear Physics Institute of the Czech Academy of Sciences, Husinec-\v{R}e\v{z}, Czech Republic\\
$^{86}$ Oak Ridge National Laboratory, Oak Ridge, Tennessee, United States\\
$^{87}$ Ohio State University, Columbus, Ohio, United States\\
$^{88}$ Physics department, Faculty of science, University of Zagreb, Zagreb, Croatia\\
$^{89}$ Physics Department, Panjab University, Chandigarh, India\\
$^{90}$ Physics Department, University of Jammu, Jammu, India\\
$^{91}$ Physics Program and International Institute for Sustainability with Knotted Chiral Meta Matter (WPI-SKCM$^{2}$), Hiroshima University, Hiroshima, Japan\\
$^{92}$ Physikalisches Institut, Eberhard-Karls-Universit\"{a}t T\"{u}bingen, T\"{u}bingen, Germany\\
$^{93}$ Physikalisches Institut, Ruprecht-Karls-Universit\"{a}t Heidelberg, Heidelberg, Germany\\
$^{94}$ Physik Department, Technische Universit\"{a}t M\"{u}nchen, Munich, Germany\\
$^{95}$ Politecnico di Bari and Sezione INFN, Bari, Italy\\
$^{96}$ Research Division and ExtreMe Matter Institute EMMI, GSI Helmholtzzentrum f\"ur Schwerionenforschung GmbH, Darmstadt, Germany\\
$^{97}$ Saga University, Saga, Japan\\
$^{98}$ Saha Institute of Nuclear Physics, Homi Bhabha National Institute, Kolkata, India\\
$^{99}$ School of Physics and Astronomy, University of Birmingham, Birmingham, United Kingdom\\
$^{100}$ Secci\'{o}n F\'{\i}sica, Departamento de Ciencias, Pontificia Universidad Cat\'{o}lica del Per\'{u}, Lima, Peru\\
$^{101}$ Stefan Meyer Institut f\"{u}r Subatomare Physik (SMI), Vienna, Austria\\
$^{102}$ SUBATECH, IMT Atlantique, Nantes Universit\'{e}, CNRS-IN2P3, Nantes, France\\
$^{103}$ Sungkyunkwan University, Suwon City, Republic of Korea\\
$^{104}$ Suranaree University of Technology, Nakhon Ratchasima, Thailand\\
$^{105}$ Technical University of Ko\v{s}ice, Ko\v{s}ice, Slovak Republic\\
$^{106}$ The Henryk Niewodniczanski Institute of Nuclear Physics, Polish Academy of Sciences, Cracow, Poland\\
$^{107}$ The University of Texas at Austin, Austin, Texas, United States\\
$^{108}$ Universidad Aut\'{o}noma de Sinaloa, Culiac\'{a}n, Mexico\\
$^{109}$ Universidade de S\~{a}o Paulo (USP), S\~{a}o Paulo, Brazil\\
$^{110}$ Universidade Estadual de Campinas (UNICAMP), Campinas, Brazil\\
$^{111}$ Universidade Federal do ABC, Santo Andre, Brazil\\
$^{112}$ Universitatea Nationala de Stiinta si Tehnologie Politehnica Bucuresti, Bucharest, Romania\\
$^{113}$ University of Derby, Derby, United Kingdom\\
$^{114}$ University of Houston, Houston, Texas, United States\\
$^{115}$ University of Jyv\"{a}skyl\"{a}, Jyv\"{a}skyl\"{a}, Finland\\
$^{116}$ University of Kansas, Lawrence, Kansas, United States\\
$^{117}$ University of Liverpool, Liverpool, United Kingdom\\
$^{118}$ University of Science and Technology of China, Hefei, China\\
$^{119}$ University of South-Eastern Norway, Kongsberg, Norway\\
$^{120}$ University of Tennessee, Knoxville, Tennessee, United States\\
$^{121}$ University of the Witwatersrand, Johannesburg, South Africa\\
$^{122}$ University of Tokyo, Tokyo, Japan\\
$^{123}$ University of Tsukuba, Tsukuba, Japan\\
$^{124}$ Universit\"{a}t M\"{u}nster, Institut f\"{u}r Kernphysik, M\"{u}nster, Germany\\
$^{125}$ Universit\'{e} Clermont Auvergne, CNRS/IN2P3, LPC, Clermont-Ferrand, France\\
$^{126}$ Universit\'{e} de Lyon, CNRS/IN2P3, Institut de Physique des 2 Infinis de Lyon, Lyon, France\\
$^{127}$ Universit\'{e} de Strasbourg, CNRS, IPHC UMR 7178, F-67000 Strasbourg, France, Strasbourg, France\\
$^{128}$ Universit\'{e} Paris-Saclay, Centre d'Etudes de Saclay (CEA), IRFU, D\'{e}partment de Physique Nucl\'{e}aire (DPhN), Saclay, France\\
$^{129}$ Universit\'{e}  Paris-Saclay, CNRS/IN2P3, IJCLab, Orsay, France\\
$^{130}$ Universit\`{a} degli Studi di Foggia, Foggia, Italy\\
$^{131}$ Universit\`{a} del Piemonte Orientale, Vercelli, Italy\\
$^{132}$ Universit\`{a} di Brescia, Brescia, Italy\\
$^{133}$ Variable Energy Cyclotron Centre, Homi Bhabha National Institute, Kolkata, India\\
$^{134}$ Warsaw University of Technology, Warsaw, Poland\\
$^{135}$ Wayne State University, Detroit, Michigan, United States\\
$^{136}$ Yale University, New Haven, Connecticut, United States\\
$^{137}$ Yildiz Technical University, Istanbul, Turkey\\
$^{138}$ Yonsei University, Seoul, Republic of Korea\\
$^{139}$ Affiliated with an institute formerly covered by a cooperation agreement with CERN\\
$^{140}$ Affiliated with an international laboratory covered by a cooperation agreement with CERN.\\

\end{flushleft} 

\end{document}